\documentclass[journal=jacsat,manuscript=article]{achemso}
\usepackage[version=3]{mhchem} 
\usepackage{xr-hyper}
\usepackage[colorlinks, linkcolor=blue,citecolor=blue, urlcolor=blue,bookmarks=true]{hyperref} 
\usepackage{import}
\usepackage[toc,page]{appendix}

\makeatletter
\newcommand*{\addFileDependency}[1]{
  \typeout{(#1)}
  \@addtofilelist{#1}
  \IfFileExists{#1}{}{\typeout{No file #1.}}
}
\makeatother

\newcounter{numsection} 
\renewcommand{\thenumsection}{S\arabic{numsection}} 

\newcommand{\numberedsection}[1]{
  \refstepcounter{numsection} 
  \section*{\thenumsection\quad #1} 
  \addcontentsline{toc}{section}{\thenumsection\quad #1} 
}

\author{Flaviano Della Pia}
\affiliation{Yusuf Hamied Department of Chemistry, University of Cambridge, Cambridge CB2 1EW, United Kingdom}
 
\author{Andrea Zen}%
\affiliation{Dipartimento di Fisica Ettore Pancini, Università di Napoli Federico II, Monte S. Angelo, I-80126 Napoli, Italy}
\alsoaffiliation{Department of Earth Sciences, University College London, London WC1E 6BT, United Kingdom}

\author{Venkat Kapil}%
\affiliation{Yusuf Hamied Department of Chemistry, University of Cambridge, Cambridge CB2 1EW, United Kingdom}

\alsoaffiliation{Department of Physics and Astronomy, University College London, London, UK}
\alsoaffiliation{Thomas Young Centre, University College London, London WC1E 6BT, United Kingdom}
\alsoaffiliation{London Centre for Nanotechnology, University College London, London WC1E 6BT, United Kingdom}

\author{Fabian L.\ Thiemann}%
\affiliation{IBM Research Europe, Keckwick Lane, Daresbury, WA4 4AD, United Kingdom}

\author{Dario Alfè}
\affiliation{Dipartimento di Fisica Ettore Pancini, Università di Napoli Federico II, Monte S. Angelo, I-80126 Napoli, Italy}
\alsoaffiliation{Department of Earth Sciences, University College London, London WC1E 6BT, United Kingdom}
\alsoaffiliation{Thomas Young Centre, University College London, London WC1E 6BT, United Kingdom}
\alsoaffiliation{London Centre for Nanotechnology, University College London, London WC1E 6BT, United Kingdom}

\author{Angelos Michaelides}
\affiliation{Yusuf Hamied Department of Chemistry, University of Cambridge, Cambridge CB2 1EW, United Kingdom}

\email{am452@cam.ac.uk, v.kapil@ucl.ac.uk}

\title{
On the increase of the melting temperature of water confined in one-dimensional nano-cavities 
}


\begin{document}

\begin{abstract}
Water confined in nanoscale cavities plays a crucial role in everyday phenomena in geology and biology, as well as technological applications at the water-energy nexus. However, even understanding the basic properties of nano-confined water is extremely challenging for theory, simulations, and experiments. In particular, determining the melting temperature of quasi-one-dimensional ice polymorphs confined in carbon nanotubes has proven to be an exceptionally difficult task, with previous experimental and classical simulations approaches report values ranging from $\sim 180 \text{ K}$ up to $\sim 450 \text{ K}$ at ambient pressure. In this work, we use a machine learning potential that delivers first principles accuracy to study the phase diagram of water for confinement diameters $ 9.5 < d < 12.5 \text{ \AA}$. We find that several distinct ice polymorphs melt in a surprisingly narrow range between $\sim 280 \text{ K}$ and $\sim 310 \text{ K}$, with a melting mechanism that depends on the nanotube diameter. These results shed new light on the melting of ice in one-dimension and have implications for the operating conditions of carbon-based filtration and desalination devices.
\end{abstract}

\section{Introduction}\label{sec:Introduction}
\noindent
Water under nanometric confinement is ubiquitous in nature and important for chemistry, physics, biology, geology, and engineering. It has received attention from both experiments and theory. Experiments suggest anomalous properties such as low dielectric response \cite{Fumagalli_low_dielectric_constant}, anomalously soft dynamics with pliable hydrogen bonds \cite{anomalous_soft_dynamics_kolesnikov}, and massive radius-dependent flow \cite{Secchi_radius_dependent_flow}. Theory and simulations indicate a quantum mechanically induced friction \cite{kakovine_quantum_friction, Bui_classical_quantum_friction}, ice-liquid oscillations \cite{molinero_oscillations}, and possible superionic behavior \cite{Monolayer_Kapil}. In addition to its potential for the discovery of new physics of confined liquids, nano-confined water has many promising applications in e.g., desalination \cite{Surwade_desalination_graphene} and clean energy \cite{Park_energy_applications,Zhang_osmotic_energy_conversion}. 
In particular, water confined in carbon nanotubes (CNTs) has been of interest for quasi one-dimensional phase transitions \cite{MD_TIP4P_Tanaka,solid_liquid_critical_behavior}, macroscopically ordered water structures \cite{dellago_long_range}, a transition from Fickian to ballistic diffusion \cite{CNT_water_diffusion}, ultra-fast water hydrodynamics \cite{science_fast_mass_transport,Shan_fast-water-transport,anomalous_soft_dynamics_kolesnikov,MD_REAXFF_Ihme}, formation of close-packed ice \cite{koga_close_packed}, as well as promising applications ranging from water purification to blue energy harvesting \cite{CNT_decontamination,CNT_membranes_w_electrochemistry_for_wastewater}.

Both experiments \cite{XRD_2005,Exp_XRD_NMR_2011_Iijima} and simulations \cite{MD_TIP4P_Tanaka,MD_REAXFF_Ihme,MD_SPCE_2007} suggest that the phase diagram of water confined in sub-nanometer tubes is significantly different from bulk water, with the formation of both ordered hollow and filled one-dimensional polymorphs, namely ice nanotubes. 
Water confined in CNTs is of strong technological interest in both solid and fluid phases.  Ice nanotubes have potential applications to ferroelectric devices \cite{ferroelectricity_1,ferroelectricity_2}, while liquid water in CNTs is important for the development of high-flux membranes \cite{highflux}, flow sensors \cite{flowsensors}, and  due to the strong analogy between CNTs and aquaporins \cite{aquaporin_1,aquaporin_2} and its potential to developing artificial water channels \cite{science_water_CNT_porins}. In this context, it is important to ask in which temperature range water confined in nanotubes melts, with implications for all the aforementioned applications.
The melting temperatures of ice nanotubes confined in CNTs have been an object of study in both simulations \cite{MD_TIP4P_Tanaka,MD_REAXFF_Ihme,MD_SPCE_2007}
and experiments \cite{XRD_2005,Exp_photoluminescence_Homma,exp_raman_extremeT_Strano}, especially below a critical confinement length scale (approximately 2.5 nm), where the macroscopic Gibbs-Thomson relation predicts a depression of
the freezing point of water, breaks down \cite{PEREZ2005709,Hugo_K_Christenson_2001}.  However, measuring the melting temperature of water in narrow carbon nanotubes is a challenge both in experiments and simulations. In fact, X-ray diffraction (XRD) measurements \cite{XRD_2005,Exp_XRD_NMR_2011_Iijima} reported melting temperatures ranging from $\sim 300 \text{ K}$ (pentagonal ice) to $\sim 180 \text{ K}$ (octagonal ice). These results are roughly in agreement with classical molecular dynamics (MD) simulations \cite{MD_TIP4P_Tanaka,MD_SPCE_2007} based on TIP4P \cite{tip4p_water} or SPC/E \cite{spce_water}
water, as well as photoluminescence (PL) experiments \cite{Exp_photoluminescence_Homma}. In contrast, Raman spectroscopy experiments \cite{exp_raman_extremeT_Strano} reported melting temperatures that were extremely sensitive to the CNT diameter, varying from $\sim 450 \text{ K}$ for $d \sim 10.5 \text{ \AA}$ to $\sim 280 \text{ K}$ for $d \sim 15.2 \text{ \AA}$. Qualitatively similar results were subsequently obtained with ReaxFF \cite{reaxff} MD simulations in Ref.\ \citenum{MD_REAXFF_Ihme}. In summary, the debate on the values of the melting temperature is still open: different experiments and (empirical force-field) simulations report values ranging from $\sim 180 \text{ K}$ up to $\sim 450 \text{ K}$.  This seemly basic disparity has large implications on the working conditions of liquid water and ice nanotubes in emerging nanotechnological applications. While different experimental techniques have been applied to investigate this problem, no computational work with the accuracy of electronic structure theory is available, mainly due to the significant length and timescale needed for reliable results. 


In this work, we take a next step towards computing the melting temperatures of one-dimensional nano-confined ice and understanding its ambient pressure phase diagram with predictive accuracy.
In particular, we achieve first-principles accuracy with feasible computational cost by using a machine learning potential (MLP) \cite{Behler_nnp,dellago_behler_mlp_water} trained on density functional theory (DFT) data, and target the question: are room temperature ice nanotubes liquid in a one-dimensionally confined CNT-like cavity?
To address this question, we study the melting temperature of nano-confined ice with an implicit model, i.e. by emulating the confining material with a cylindrical confining potential fitted to the water-carbon interaction in sub-nanometer carbon nanotubes.
This is a standard approach in analyzing the phase behavior of quasi one-dimensional nano-confined water \cite{Koga_ordered_ice_nanotubes,tip4p_water,Exp_XRD_NMR_2011_Iijima}, and we have checked the reliability of our implicit model towards the explicit modeling of the carbon atoms by using the MLP developed in Ref.\ \citenum{Thiemann_Schran_water_flow}.
To compute the melting temperatures, we determine the most stable polymorphs for a fixed nanotube diameter by using random structure search (RSS) \cite{Pickard_2011} and compute its melting temperature via solid-liquid coexistence simulations. We find the melting points of (helical) triangular, square, pentagonal, and hexagonal ice nanotubes to be $\sim 10-30 \text{ K}$ higher than the bulk water melting temperature. In addition, we report a non monotonic behavior of the number of hydrogen bonds with the confining diameter, that positively correlates with the water diffusion coefficient. 
On the one hand, our results confirm the possibility to study and apply ice nanotubes at around room temperature, but suggest that the range of stability is limited at temperatures below $ \sim 310 \text{ K}$, as opposed to a previously higher suggested range. On the other hand, our results indicate that filtration and desalination devices based on water confined in narrow tubes do not require high working temperatures.

\section*{The melting temperature of quasi one-dimensional nano-confined ice}\label{sec:Results}
\noindent
Previous experimental and empirical force-field based computational work suggests that water confined in sub-nanometer nanotubes exhibits an interesting phase diagram, with different quasi one-dimensional ice polymorphs stable in different diameter ranges \cite{MD_TIP4P_Tanaka,Exp_photoluminescence_Homma,Exp_XRD_NMR_2011_Iijima}. In this work, we explore the phase behavior of ice confined in narrow nanotubes, i.e. with a diameter $d \sim 10 \text{ \AA}$, at first principles accuracy. Therefore, we considered five confining cylinders corresponding to zigzag nanotubes CNT(n,0) with $n=12,13,14,15,16$. The diameters of the considered nanotubes are respectively $\sim 9.5, 10.2, 11.0, 11.8, 12.5 \text{ \AA}$. The diameters were selected to maximize the variability in the studied phase diagram. In fact, a different one-dimensional ice polymorph is expected to be the most stable polymorph for each confining diameter \cite{MD_TIP4P_Tanaka}.

To explore the phase behavior of the ice nanotubes, we use an implicit confinement model. We first developed a uniform cylindrical confining potential fitted to revPBE-D3 \cite{revPBE,Grimme_D3_1} binding energies of a water molecule inside the CNT. The DFT functional for the water-carbon interaction was determined via an accurate benchmark to diffusion Monte Carlo (DMC) and Coupled Cluster with Single, Double, and perturbative Triple interactions (CCSD(T)) data from Ref.\  \citenum{Brandenburg_water_carbon_2019}. The water-water interaction is described by using the MLP from Ref.\  \citenum{Pavan_monolayer}, trained on revPBE0-D3 \cite{revPBE0,Grimme_D3_1} for treating water in bulk and under confinement \cite{Monolayer_Kapil,Pavan_monolayer}. To find all the metastable phases in each nanotube, we performed an RSS with a home-built Python code. Subsequently, we identified the most stable polymorph in each nanotube based on the minimization of the enthalpy, and then computed its melting temperature via coexistence simulations. The reliability of our model in determining the lowest enthalpy structure has been checked against both revPBE0-D3 and DMC data, as reported in the Supporting Information (SI)\cite{supporting_information} (section~\ref{si-sec:dmc}).  Further technical details on the RSS, the fitting of the confining potential, and the coexistence simulations are reported in \textbf{Methods} and in the SI, together with tests on the robustness of our model with respect to the modeling of explicit carbon (section~\ref{si-sec:explicit}). Considering the weak impact of quantum nuclear motion on the melting temperature of bulk \cite{Cheng_PNAS_water} and 2D nanoconfined water \cite{Monolayer_Kapil} at ambient pressures, we restrict ourselves to a classical description of the nuclei.

\begin{figure*}[tbh!]
    \centering
    \includegraphics[scale=0.8]{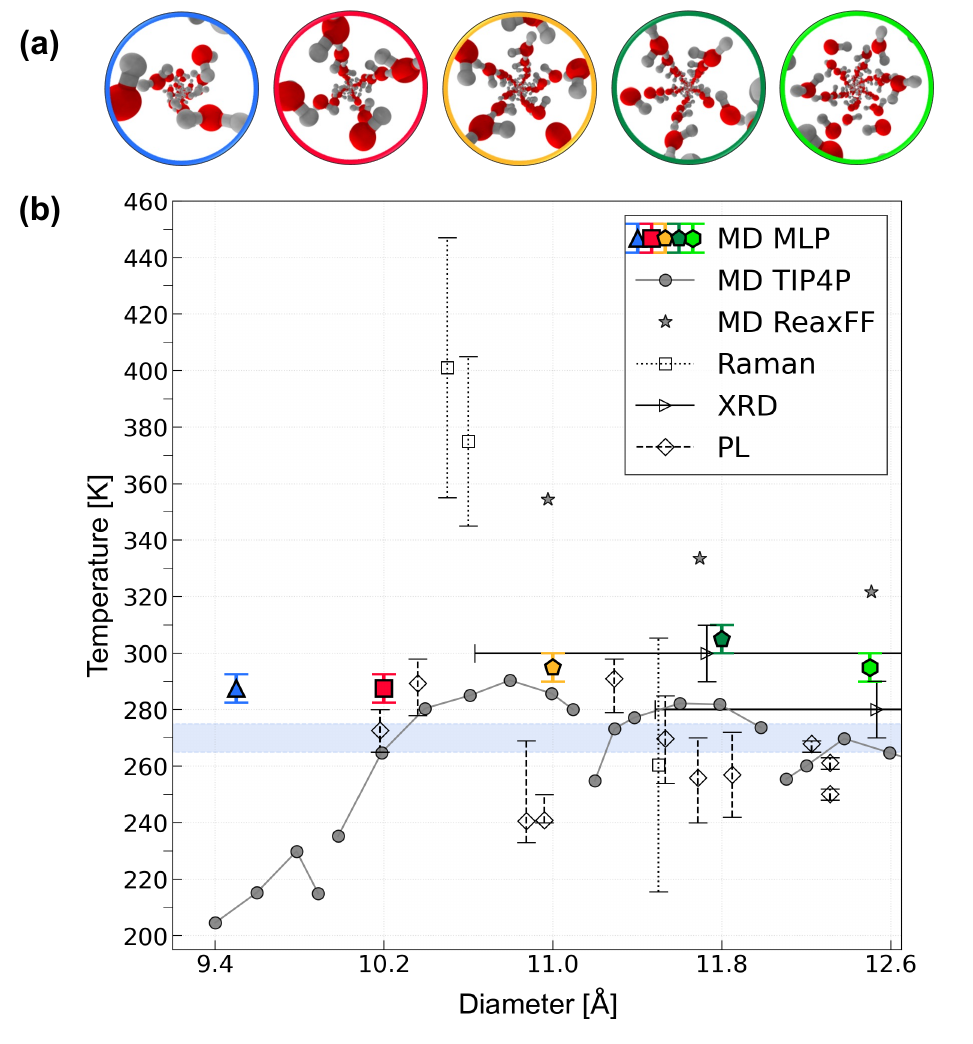}
    \caption{ Melting temperature of nano-confined ice tubes. (a) Snapshots of the solid structures stable at zero temperature. Oxygen atoms are plotted in red and hydrogen atoms are plotted in gray. (b) The melting temperature of triangular (blue), square (red), pentagonal (orange and dark green), and hexagonal (light green) ice computed in this work are reported as a function of the confinement diameter. Empty black markers show experimental results. In particular we use diamonds with dashed error bars for PL \cite{Exp_photoluminescence_Homma}, triangles with solid error bars for XRD \cite{XRD_2005,Exp_XRD_NMR_2011_Iijima}, and squares with dotted error bars for Raman spectroscopy \cite{exp_raman_extremeT_Strano}.
    Previous empirical force-field MD simulations are reported with gray stars for results from Ref.\  \citenum{MD_REAXFF_Ihme} obtained with ReaxFF, and gray circles for results from Ref.\  \citenum{MD_TIP4P_Tanaka} obtained with TIP4P. Different curves in the TIP4P results refer to ranges where a different ice nanotube is stable.
    The melting temperature of bulk water according to our MLP is $270 \pm 5 \text{ K}$ \cite{Monolayer_Kapil}, and is shown on the plot with a shaded light blue area.}
    \label{fig:melting}
\end{figure*}

The most stable phases identified with our approach in the five considered nanotubes are (helical) triangular ($d\sim 9.5 \text{ \AA}$), square ($d\sim 10.2 \text{ \AA}$), pentagonal ($d\sim 11.0$ and $\sim 11.8 \text{ \AA}$), and hexagonal ice ($d\sim 12.5 \text{ \AA}$). Front-view snapshots of the zero-temperature structures are reported in Fig.\ \ref{fig:melting}(a).
We refer to the melting temperature - diameter phase diagram in Fig.\ \ref{fig:melting}(b). The first principles accuracy melting points obtained in this work are reported with a blue triangle, red square, orange and dark green pentagon, and light green hexagon.
 We find that the melting temperatures of ice nanotubes are $\sim 10-30 \text{ K}$ above the melting temperature of bulk water, which is $270 \pm 5 \text{ K}$ for our model \cite{Monolayer_Kapil}. In particular, the melting point of the square ice nanotube is in quantitative agreement with PL \cite{Exp_photoluminescence_Homma} spectroscopy. A much higher transition temperature was reported with Raman spectroscopy at similar diameters \cite{exp_raman_extremeT_Strano}. However, Chiasci \textit{et al.} \cite{Exp_photoluminescence_Homma} argue that the high temperature reported in Ref.\  \citenum{exp_raman_extremeT_Strano} could be related to the observation of the encapsulation process (the vapor-liquid phase transition). The melting temperature of pentagonal and hexagonal ice nanotubes are in near quantitative agreement with XRD \cite{XRD_2005,Exp_XRD_NMR_2011_Iijima} and Raman \cite{exp_raman_extremeT_Strano} experiments, considering the large experimental error bars and uncertainties on the confining diameter.
 
It is not straightforward to compare our results to the melting temperatures predicted by empirical water models. In fact, the MLP melting temperatures of square, pentagonal, and hexagonal ice are approximately $20-30 \text{ K}$ higher than TIP4P, while the MLP melting temperature of triangular ice is $\sim 90 \text{ K}$ higher than the TIP4P prediction. However, TIP4P predicts a bulk melting temperature of $\sim 230 \text{ K}$ \cite{water_ff_melting}. Hence, TIP4P predicts a melting temperature of triangular ice approximately $20 \text{ K}$ lower than the bulk, while the TIP4P predicted melting temperatures of square, pentagonal, and hexagonal ice are approximately $30-50 \text{ K}$ higher than the bulk. 
 In general, the melting point of water is very sensitive to the force-field used to describe the water-water interaction, both in bulk \cite{water_ff_melting} and under confinement \cite{JCP_2d_HSTI,2d_ff_order_of_phase_transition}. Ref. \citenum{2d_ff_order_of_phase_transition} shows for instance a discrepancy of $300 \text{ K}$ across empirical forcefields in the predicted melting temperature for 2D water. In addition, we show in the SI (section~\ref{si-sec:distances_angles}) that the spectra of bond lengths and angles in the optimized ice nanotubes differ from the fixed value considered in rigid models, a result similar to that found in 2D confinement \cite{Monolayer_Kapil}. Overall this analysis emphasizes the need for achieving predictive ability with first principles accuracy, and suggests that our work provides valuable insight into the phase diagram of nano-confined ice.

\section*{Continuous or discontinuous phase transition?}
\noindent The nature of the melting phase transition is of special interest for water under confinement. In fact, while the melting in bulk systems is a first-order direct process, in low dimensional systems it can be more complex. For instance, 2D ice has been predicted to melt into a liquid via a hexatic phase \cite{hexatic_2d_artacho,Monolayer_Kapil}. The order of phase transitions resembles a first order for solid to hexatic, but second order for hexatic to liquid \cite{Monolayer_Kapil}. 
In the case of CNTs, previous empirical force-field MD studies showed that water in carbon nanotubes may freeze either continuously or discontinuously, with strong sensitivity on diameter and pressure \cite{MD_TIP4P_Tanaka,Koga_ordered_ice_nanotubes,MD_REAXFF_Ihme}. In particular, at ambient pressure TIP4P simulations from Ref.\ \citenum{MD_TIP4P_Tanaka} suggest that the melting transition is continuous for pores with diameters $d > 12 \text{ \AA}$, while it can be both continuous or discontinuous for $d< 12 \text{ \AA}$. In contrast, ReaxFF simulations from Ref.\ \citenum{MD_REAXFF_Ihme} suggest that the phase transition is discontinuous for hexagonal ice, continuous for square and pentagonal ice, while triangular ice undergoes a supercritical transition with the absence of a diffusive regime at high temperatures. Indeed, it has been shown for 2D nano-capillaries that the order of solid-liquid phase transitions of empirical forcefields is sensitive to their parametrization \cite{2d_ff_order_of_phase_transition}.

\begin{figure*}[tbh!]
    \centering
    \includegraphics[scale=0.7]{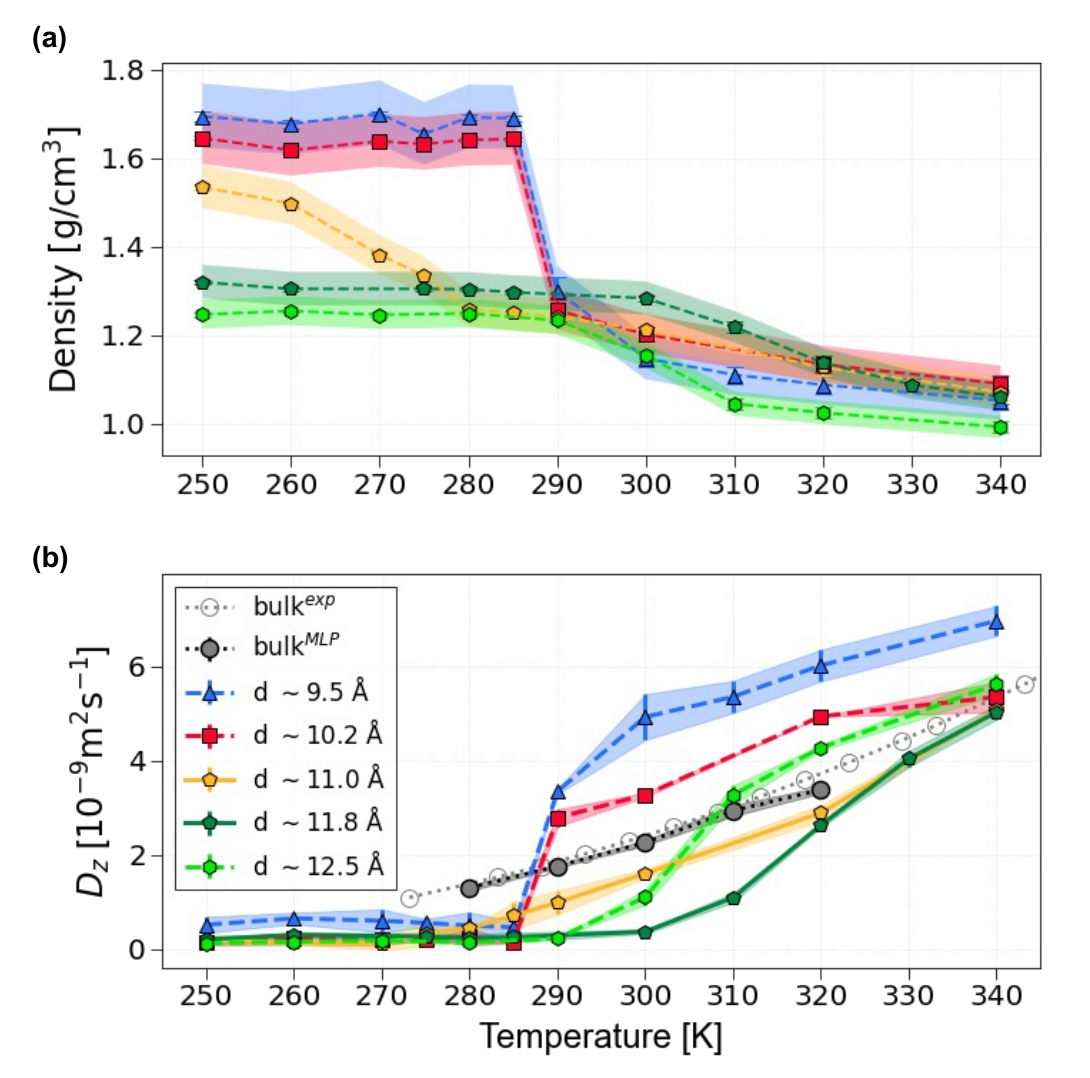}
    \caption{Melting phase transition of nano-confined ice tubes. We report the density (panel a) and the diffusion coefficient $D_z$ along the cylinder axis (panel b) as a function of temperature. In panel (b), we also report the diffusion coefficient of bulk water computed with our MLP and compared to experimental results \cite{exp_bulk_diffusion_Sacco}. Both the density and the diffusion coefficient show signs of a discontinuous phase transition for triangular and square ice, and a continuous phase transition for pentagonal and hexagonal ice.}
    \label{fig:phase_transition}
\end{figure*}

In Fig.\ \ref{fig:phase_transition}, we report the density (a) and the diffusion coefficient $D_z$ along the nanotube axis (b) as a function of temperature. We also report the diffusion coefficient of bulk water computed with the MLP and compared to experimental results \cite{exp_bulk_diffusion_Sacco}, showing the accuracy of our model with quantitative agreement from $280 \text{ K}$ to $320 \text{ K}$. Details on the calculation of density and the diffusion coefficient are given in \textbf{Methods} and in the SI (sections~\ref{si-sec:coexistence-simulations} and \ref{si-sec:diffusion-coefficient}). 
With our model, both structure (density) and dynamics (diffusion) suggest that the phase transition can be either continuous or discontinuous. In particular, we observe hallmarks of a discontinuous phase transition for triangular and square ice (stronger confinement regime). On the other hand, seemingly smooth changes in the density and the diffusion coefficient indicate a continuous phase transition for pentagonal and hexagonal ice. As the melting temperature and the order of phase transitions are sensitive to finite size effects, we show in the SI (section \ref{si-sec:fse}) that our results are converged with respect to the system size. Overall, such contrasting melting behavior observed within such a narrow range of diameters is a clear illustration of the delicate and fascinating behavior of nano-confined water.

\section*{Structure and dynamics of nano-confined water depend non-monotonically on the nanotube diameter}
\begin{figure*}
    \centering
    \includegraphics[scale=0.7]{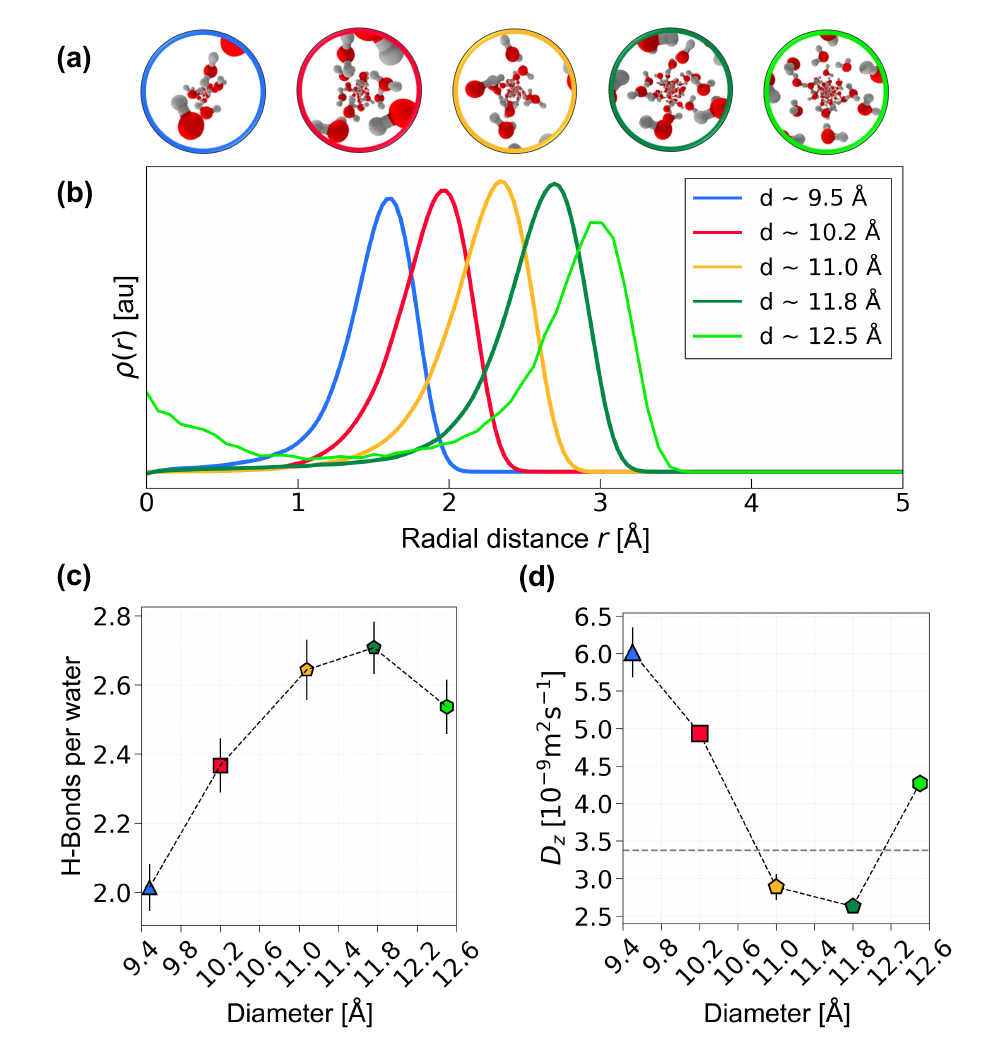}
    \caption{Structural and dynamical properties of liquid water for different confining diameters at $T=320\text{ K}$. (a) Snapshots of the liquid structures. (b) Radial density $\rho(r)$ as a function of the radial distance $r$. (c) Average number of hydrogen bonds per water molecule as a function of the nanotube diameter at room temperature. (d) Diffusion coefficient $D_z$ along the nanotube axis as a function of the diameter for $T=320 \text{ K}$. The bulk value at the same temperature computed with our model is reported with a black dashed line. Increasing the confinement to a diameter $d < 12 \text{ \AA}$ defines a more centred liquid, with a decreasing number of hydrogen bonds and a corresponding increase in the diffusion coefficient.}
    \label{fig:structure-dynamics}
\end{figure*}

\noindent 
So far we focused on identifying the melting temperatures of nano-confined ice tubes. To gain further insight into these systems and to try to understand our observations, we now look at the properties of the liquid as a function of the confinement diameter.
In Fig.\ \ref{fig:structure-dynamics}, we report an analysis on structure and dynamics of the liquid equilibrated at the temperature $T = 320 \text{ K}$. In particular, we compute the density as a function of the radial distance in the confining cylinder, the average number of hydrogen bonds via geometrical criteria defined in Ref.\ \citenum{hb_chandler}, and the water diffusion coefficient using the velocity autocorrelation function (VACF) method \cite{frenkel2023understanding}. We observe a non monotonic behavior of the average number of hydrogen bonds per molecule as a function of the confining diameter, that correlates positively with the non monotonic trend in the diffusion coefficient.

Increasing the diameter from $9.5 \text{ \AA}$ to $11.8 \text{ \AA}$ defines a less centred liquid, with the peak of the radial density as a function of the distance from the centre $r$ (panel b) shifting from $r \sim 1.5  \text{ \AA}$ to $r \sim 2.6  \text{ \AA}$. This change is accompanied by an increase in the average number of hydrogen bonds (panel c) from $\sim 2.0$ to $\sim 2.7$. The number of hydrogen bonds in the smallest diameter is consistent with both our RSS and previous results \cite{MD_TIP4P_Tanaka}, because only centred chains of water molecules are expected to be stable for $d < 9 \text{ \AA}$.
The liquid structure in the largest diameter ($12.5 \text{ \AA}$) consists of a chain of water molecules inside water rings, in agreement with previous observations \cite{Koga_ordered_ice_nanotubes,MD_TIP4P_Tanaka,CNT_water_diffusion} . This results in a peak in the radial density close to $r \sim 0$ and correlates with a decrease in the number of hydrogen bonds.

The liquid phase diffusion coefficient $D_z$ (panel d) shows a consistent trend with the number of hydrogen bonds, increasing from $\sim 3.0 \times 10^{-9}\text{m}^2\text{s}^{-1}$ at $11.8 \text{ \AA}$ to $\sim 5.5 \times10^{-9}\text{ m}^2\text{s}^{-1}$ at $9.4 \text{ \AA}$. In particular, the diffusion coefficient for $d< 10 \text{ \AA}$ is higher than the bulk value at the same temperature, which is $\sim 3.8 \times 10^{-9} \text{ m}^2\text{s}^{-1}$ with our model. The increase of the diffusion coefficient with stronger confinement is consistent with recent experiments \cite{Shan_fast-water-transport} reporting ultra-fast transport (slip length of $\sim 8.5 \text{ µm}$) in vertically aligned CNTs membrane with $d < 9 \text{ \AA}$. Qualitatively similar behavior was found with TIP4P/2005 and SPC/E, but not with other empirical force-field simulations, predicting a diffusion coefficient increasing with an increasing diameter in the sub-nanometer regime \cite{diffusion_ff_Chacham,properties_water_cnt_Hassan, MD_REAXFF_Ihme,CNT_water_diffusion}.

In summary, we observe that a different liquid is associated with a different confining diameter and ice nanotube, with the melting temperatures varying in a relatively narrow range of $\sim 30 \text{ K}$. As the melting temperature depends on the relative Gibbs free energy between a solid and the liquid, we think that the seemingly weak radius dependence of the melting temperature could arise from an overall cancellation of the effects of confinement in both the solid and liquid phases. Finally, we acknowledge that the exact values of both structural and dynamical properties of confined water are expected to be influenced by the modeling of the confining material. However, in the SI (section~$\ref{si-sec:explicit}$) we show that our results are consistent with respect to the modeling of explicit CNTs. In fact, we report on the diffusion coefficient and the average number of hydrogen bonds in the three largest nanotubes considered in this work and show results that are in qualitative agreement with those reported in the main manuscript with the uniform confining potential.

\section*{Conclusions}\label{sec:Conclusions}

\noindent In this work, we computed the melting temperatures of quasi one-dimensional ice nanotubes with first principles accuracy.
This topic has been debated in the last decades, with both experiments and classical simulations reporting melting temperatures ranging from $\sim 180 \text{ K}$ to $\sim 450 \text{ K}$. Exploiting machine learning based first principles accuracy simulations, we show that one-dimensional nano-confined water is liquid around room temperature. In particular,  
we computed the melting temperature of (helical) triangular, square, pentagonal, and hexagonal ice in a cylindrical confining potential, respectively of diameter $ d \sim 9.5, 10.2, 11.0, 11.8,$ and $ 12.5 \text{ \AA}$. We find that all the considered ice nanotubes melt in the range $\sim 280-310 \text{ K}$.  
Our melting temperature of square ice is in agreement with PL spectroscopy experiments \cite{Exp_photoluminescence_Homma}; similarly, the melting temperatures of pentagonal and hexagonal ice are in agreement with Raman, XRD, and PL spectroscopy experiments \cite{exp_raman_extremeT_Strano,XRD_2005,Exp_photoluminescence_Homma,Exp_XRD_NMR_2011_Iijima}. In addition, we report our melting temperature of triangular ice is predicted to be higher than previously reported. Notably, empirical force-field simulations predicted melting temperatures differing by as much as $\sim 80 \text{ K}$ from our MLP, as well as qualitatively different behavior in the diffusion coefficient. 
Finally, we provide an insight into the structural and dynamical properties of water confined in different confinement regimes. In a sub-nanoscale confinement regime ($d < 10 \text{ \AA}$), we report a strong reduction in the number of hydrogen bonds and a corresponding enhanced diffusion coefficient that exceeds the bulk limit. 

Our model is certainly limited in capturing the effects of chirality and phonon-coupling of the CNTs, due to the use of a uniform confining potential. However, we show that the uniform confining potential remains a good model for studying the phase behavior of quasi 1D water. In addition, by changing the confining potential our simplified model could easily be adapted to the study of different confinement configurations, such as 2D slits of different widths, nanotubes of different diameters, or conical confinement configurations.

In summary, this work suggests that first principles accuracy is required for the modeling of one-dimensional nano-confined water. In addition, it improves on the understanding of the melting transition, structural and dynamical properties of nano-confined water, providing further insight for technological application for water filtration and desalination, or in the development of artificial channels mimicking biological systems.

\section{Methods}\label{sec:Methods}
\noindent
To ensure both computational efficiency and accuracy, we follow a similar approach used in Ref.\ \citenum{Monolayer_Kapil}. We used: (1) a combination of available DMC and CCSD(T) data to select a DFT functional to describe the water-carbon interaction and parametrise a Morse potential; (2) an MLP trained on bulk and confined water structures for the water-water interaction \cite{Monolayer_Kapil,Pavan_monolayer}; (3) random structure search (RSS) to identify metastable phases; and (4) solid-liquid coexistence simulations to compute the melting temperatures. 

\section*{Separation of the potential energy surface}
\noindent
We split the potential energy of the system into: (1) water-CNT interaction, modeled using a radial confining potential fitted to DFT water-CNT binding energies; and (2) water-water interactions described by an MLP trained on DFT data.
The water-CNT interaction is modeled with the revPBE-D3 functional, selected according to the benchmark on DMC and CCSD(T) (previously computed in Ref.\  \citenum{Brandenburg_water_carbon_2019}) data reported in the SI (section~\ref{si-sec:benchmark}).
The water-water interaction is modeled with the MLP trained on revPBE0-D3 data for bulk and confined water in Refs.\  \citenum{Monolayer_Kapil,Pavan_monolayer}. The MLP model relies on the Behler–Parrinello neural network framework \cite{Behler_nnp} to form a committee neural network potential \cite{schran_committee} and trained with an active learning framework \cite{schran_pnas}.
The DFT functional was selected based on DMC benchmarks \cite{Monolayer_Kapil}, and the model has been already applied to the analysis of the phase diagram of monolayer confined water \cite{Monolayer_Kapil,Pavan_monolayer}    .

\section*{Structure search}
\noindent
We probe potential phases of ice nanotubes using the RSS approach in combination with the MLP. Within this approach, we recover the previously known ice polymorphs by optimizing a large set of structures generated by randomly placing water molecules within the confinement region at ambient pressure. The RSS is less suitable for the identification of helical structures, which require a specific number of molecules. We build helical ice nanotubes according to the theory described in Ref.\ \citenum{MD_TIP4P_Tanaka} and optimize them with our model. The zero-temperature-zero-pressure stable phase is subsequently identified as the one with the lowest energy. The validity of our model has been tested towards DMC data for the CNT(15,0), as described in the SI.

\section*{Coexistence simulations}
\noindent
The melting temperatures of the confined ice nanotubes are determined via solid-liquid coexistence simulations.
An initial solid structure is thermalized at $\sim 250 \text{ K}$; half of the oxygens are subsequently frozen, while the other half is melted at $\sim 600 \text{ K}$ and then quenched down to $\sim 300 \text{ K}$. The interface between the solid and the liquid is built in the NVT ensemble to avoid large box fluctuations during the high-temperature melting phase.
The coexistence simulations are subsequently run in the $NP_z T$ ensemble with $P \sim 10 \text{ bar}$ and changing temperatures in the range $\sim [250,340]  \text{ K}$.

The melting temperature is determined according to changes in the density and the diffusion coefficient as a function of temperature, as shown in Fig.\ \ref{fig:phase_transition}.
The density is computed as the ratio of the number of molecules and the occupied volume inside the confining cylinder. To define the volume of the cylinder occupied by water molecules, we consider the radial density $\rho(r)$ as a function of the radial distance $r$. We define $r_{\text{max}}$ as the maximum distance $r$ such that $\rho(r)>0$. The occupied volume $V$ is computed as $V=\pi r^2_{\text{max}} l_{\text{z}}$, where $l_{\text{z}}$ is the length of the simulation box. The shaded error bars in Fig.\ \ref{fig:phase_transition}(a) are computed considering a $\sim 5\%$ uncertainty in the definition of $r_{\text{max}}$.
The diffusion coefficient of nano-confined ice is estimated using the VACF method. A comparative analysis on the estimate of the diffusion coefficient obtained by using the Einstein relation to extract the diffusion coefficient from the Mean Square Displacement (MSD)\cite{frenkel2023understanding} and VACF is reported in the SI. The two approaches deliver results in close agreement, and the VACF approach was chosen to present results in the main manuscript as it delivers smaller sampling uncertainties.
The diffusion coefficient of the bulk is computed with the MSD method using a cubic box of side $L=24.84 \text{ \AA}$ containing $512$ water molecules, and applying the temperature-dependent finite-size correction from Ref.\ \citenum{diffusion_fse_bulk}.

\section*{Computational details}
\noindent
Molecular Dynamics (MD) simulations are performed using the i-PI \cite{i-pi} code with the n2p2-LAMMPS \cite{n2p2_lammps,LAMMPS} library to calculate the MLP energies and forces, and an ASE \cite{ase} driver for the radial confining potential. The time-step is fixed to $0.5 \text{ fs}$, and pressure and temperature are controlled with the generalized Langevin equation (GLE) barostat-thermostat as implemented in i-PI. Complete details on the development of the used MLP are given in Ref.\ \citenum{Pavan_monolayer}.

Coexistence simulations have been performed with supercells containing $\sim 700$ water molecules ($\sim 40 \text{ nm}$ long nanotubes) to limit finite-size effects. Tests on the effect of the number of water molecules on the density and the diffusion coefficient are reported in the SI (section~\ref{si-sec:fse}). In particular, we show that density in our simulations is converged compared to simulations with $\sim 10^4$ water molecules, which is comparable with the number of molecules expected in experiments with CNTs of length $\sim 1 \mathrm{ \mu m}$. The length of the MD simulations varies from $10-20 \text{ ns}$ depending on when convergence is achieved, as shown in the SI.

The water-CNT confining potential is fitted to revPBE-D3 binding energies computed with a $1\times 1 \times 3$ k-point grid and $\sim 9 \text{ \AA}$ long nanotubes. All calculations are run using VASP \cite{VASP1,VASP2,VASP3,VASP4} with a $1000 \text{ eV}$ energy cut-off, fine FFT grids and hard pseudopotentials. Further details on the construction of the confining potential are given in the SI (section~\ref{si-sec:confining-potential}).

The DMC calculations are performed using the CASINO \cite{CASINO} package, using eCEPP \cite{CASINO_PSEUDO_eCEPP} pseudopotentials with the determinant-locality approximations \cite{ZenDLA}, and taking into account errors due to finite system size \cite{MPC1,MPC2,MPC3} and finite time-step \cite{ZSGMA} (convergence shown in the SI). This setup was also used to study bulk ice \cite{DMCICE13}, yielding results in excellent agreement with experiments.

\begin{suppinfo}
\noindent
See the Supporting Information for details on the CNTs confining potential, validation of our model with diffusion Monte Carlo, the analysis of the coexistence simulations, tests on the estimates of the diffusion coefficient, and comparison between predictions of the implicit and explicit carbon models.
\end{suppinfo}

\begin{acknowledgement}
We thank all members of the ICE group for valuable feedback on early stages of the manuscript. 
We acknowledge the computational resources from Cambridge Service for Data Driven Discovery (CSD3) operated by the University of Cambridge Research Computing Service (www.csd3.cam.ac.uk), provided by Dell EMC and Intel using Tier-2 funding from the Engineering and Physical Sciences Research Council (capital grant EP/T022159/1 and EP/P020259/1), and DiRAC funding from the Science and Technology Facilities Council (www.dirac.ac.uk). 
We are furthere grateful for computational support from the UK national high performance computing service, ARCHER2, for which access was obtained via the UKCP consortium and funded by EPSRC grant ref EP/X035891/1, and Swiss National Supercomputing Centre under project s1209.
V.K. acknowledges support from the Ernest Oppenheimer Early Career Fellowship and the Sydney Harvey Junior Research Fellowship.
A.M. acknowledges support from the European Union under the “n-AQUA” European Research
Council project (Grant No. 101071937). 
D.A. and A.Z. acknowledges support from Leverhulme grant no. RPG-2020-038, and from the European Union under the Next generation EU (projects 20222FXZ33 and P2022MC742).
\end{acknowledgement}

\bibliography{ref}

\begin{appendices}

In the supporting information we provide:
\begin{itemize}
    \item the benchmark of 42 density functional theory (DFT) functionals for the water-carbon interaction in section \ref{si-sec:benchmark};
    \item the confining potentials for the carbon nanotubes (CNTs) with revPBE-D3 in section \ref{si-sec:confining-potential};
    \item the validation of our model with diffusion Monte Carlo (DMC) in section \ref{si-sec:dmc};
    \item the analysis on the convergence of our coexistence simulations in section \ref{si-sec:coexistence-simulations};
    \item the analysis on the estimates of the diffusion coefficient in section \ref{si-sec:diffusion-coefficient};
    \item the comparison of our (implicit) model with an explicit model containing carbon atoms in section \ref{si-sec:explicit};
    \item the analysis of the finite size error in our coexistence simulations in section \ref{si-sec:fse};
    \item the analysis of the bonds and angles distributions in the optimized ice nanotubes in section \ref{si-sec:distances_angles}.

\end{itemize}

\numberedsection{Benchmark of (42) DFT functionals for water-carbon interaction}\label{si-sec:benchmark}
In the main manuscript, we describe the water-wall interaction with a confining potential fitted to DFT data of the binding energy of a water molecule inside the CNT.
The first step towards developing such a confining potential is determining an accurate functional for the water-carbon interaction. This problem has been previously addressed in Ref.\ \citenum{Brandenburg_water_carbon_2019}, where 28 DFT functionals were benchmarked to high-accuracy computational reference values. 
Here, we extend on Ref.\ \citenum{Brandenburg_water_carbon_2019} by testing 31 additional functionals. The benchmark consists in computing the binding energies of a single water molecule to benzene, coronene and graphene (with three possible orientations of the water molecule), and both inside and outside the CNT(10,0). Reference values from Ref.\ \citenum{Brandenburg_water_carbon_2019} were computed either with DMC or coupled cluster with single, double and perturbative triple excitations [CCSD(T)]. 

The DFT calculations are performed using VASP\cite{VASP1,VASP2,VASP3,VASP4}. The binding energies were computed at the $\Gamma$ point for benzene, coronene and the CNT(10,0). A $5\times 5\times 1$ k-point grid was used for graphene (with a $50$ atoms supercell), except for meta-GGA and hybrid functionals computed at $\Gamma$. Convergence of the computational set-up has been tested in Ref.\ \citenum{Brandenburg_water_carbon_2019}. Geometries were taken from Ref.\ \citenum{Brandenburg_water_carbon_2019}.

\begin{figure}[tbh!]
    \centering
    \includegraphics[scale=0.8]{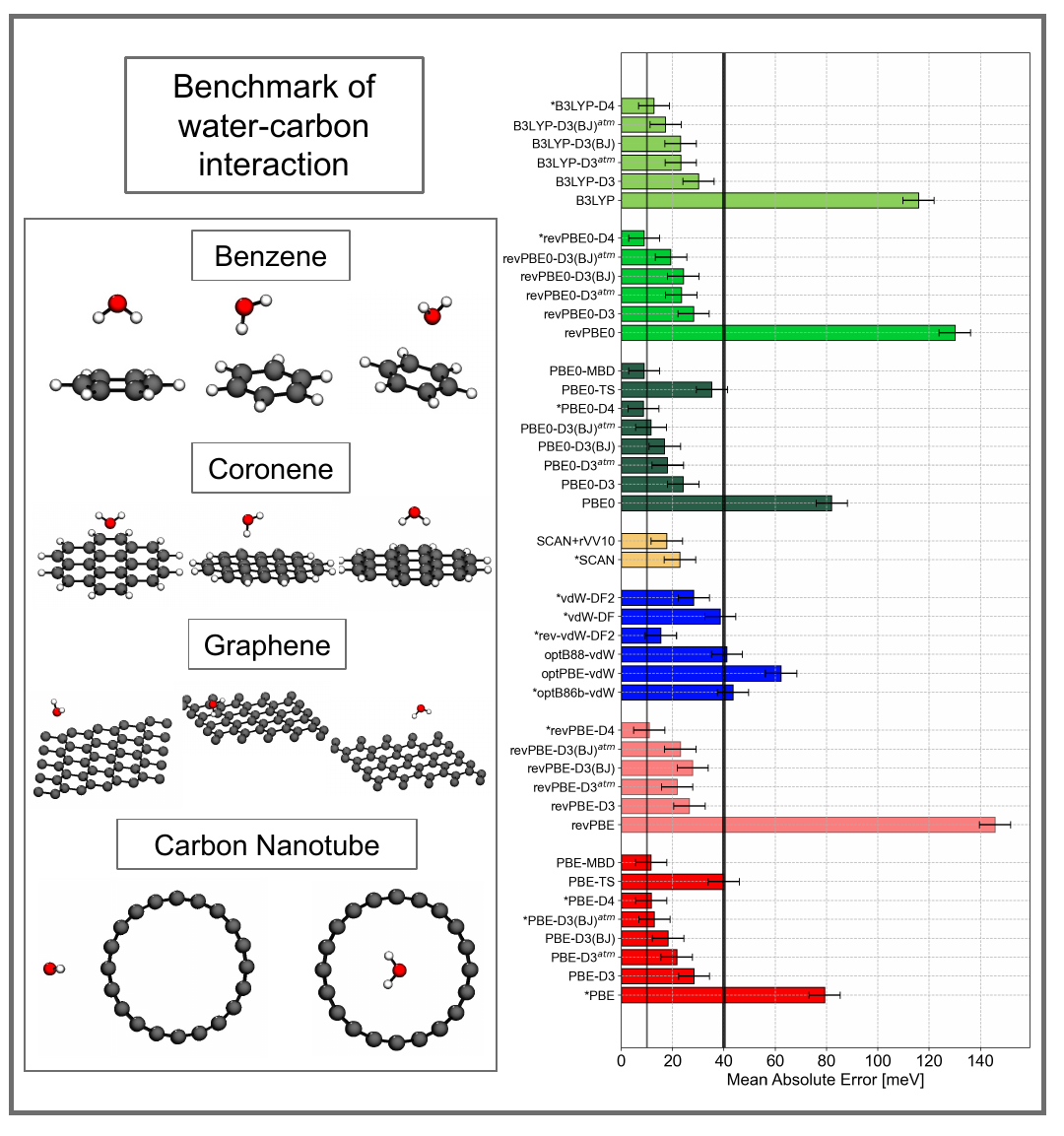}
    \caption{Benchmark of DFT functionals for water-carbon interaction. The benchmark is conducted on binding energies of a single water molecule to benzene, coronene, graphene, and inside and outside the CNT(10,0). The performance of each functional is reported as a MAE with respect to reference values from Ref.\ \citenum{Brandenburg_water_carbon_2019}. The error bars are computed as the average error of the reference values. Grey vertical lines refer to the chemical ($40$ meV) and sub-chemical ($40$ meV) accuracy. The 11 functionals that have been previously tested\cite{Brandenburg_water_carbon_2019}
    are indicated with an asterisk.}
    \label{fig:si-benchmark}
\end{figure}

In Fig.\ \ref{fig:si-benchmark} we report the performance of each tested functional as a Mean Absolute Error (MAE) with respect to the reference values. The 11 functionals that have been previously tested are indicated with an asterisk. Grey vertical lines highlight the chemical accuracy ($40$ meV) and a sub-chemical accuracy limit of $10$ meV.
Overall, several functionals achieve chemical or sub-chemical accuracy. Considering that revPBE-D3: (1) has a reliable performance for both water-carbon interaction and water-water in ice polymorphs \cite{DMCICE13}; (2) it is the functional used in the explicit water-CNT model tested afterwards\cite{Thiemann_Schran_water_flow}, this was chosen to compute the confining potential.

\clearpage


\clearpage
\numberedsection{Single water molecule in carbon nanotubes: confining potentials}\label{si-sec:confining-potential}
The confining potentials are computed by fitting the binding energy of a single water molecule inside the CNT to a Morse potential. 
We computed the energy as a function of the radial position of the water molecule averaging along 12 possible directions ($\pm 0$, $\pm 30\text{°}$, $\pm45\text{°}$, $\pm60\text{°}$,$\pm90\text{°}$).
The confining potentials are functions of the distance from the wall, therefore neglect the surface roughness of the nanotube. Tests (moving the water molecule along the CNT axis) showed that the corrugation changes the potential of $<20 \text{ meV}$.

\begin{figure}[tbh!]
    \centering
    \includegraphics[scale=0.8]{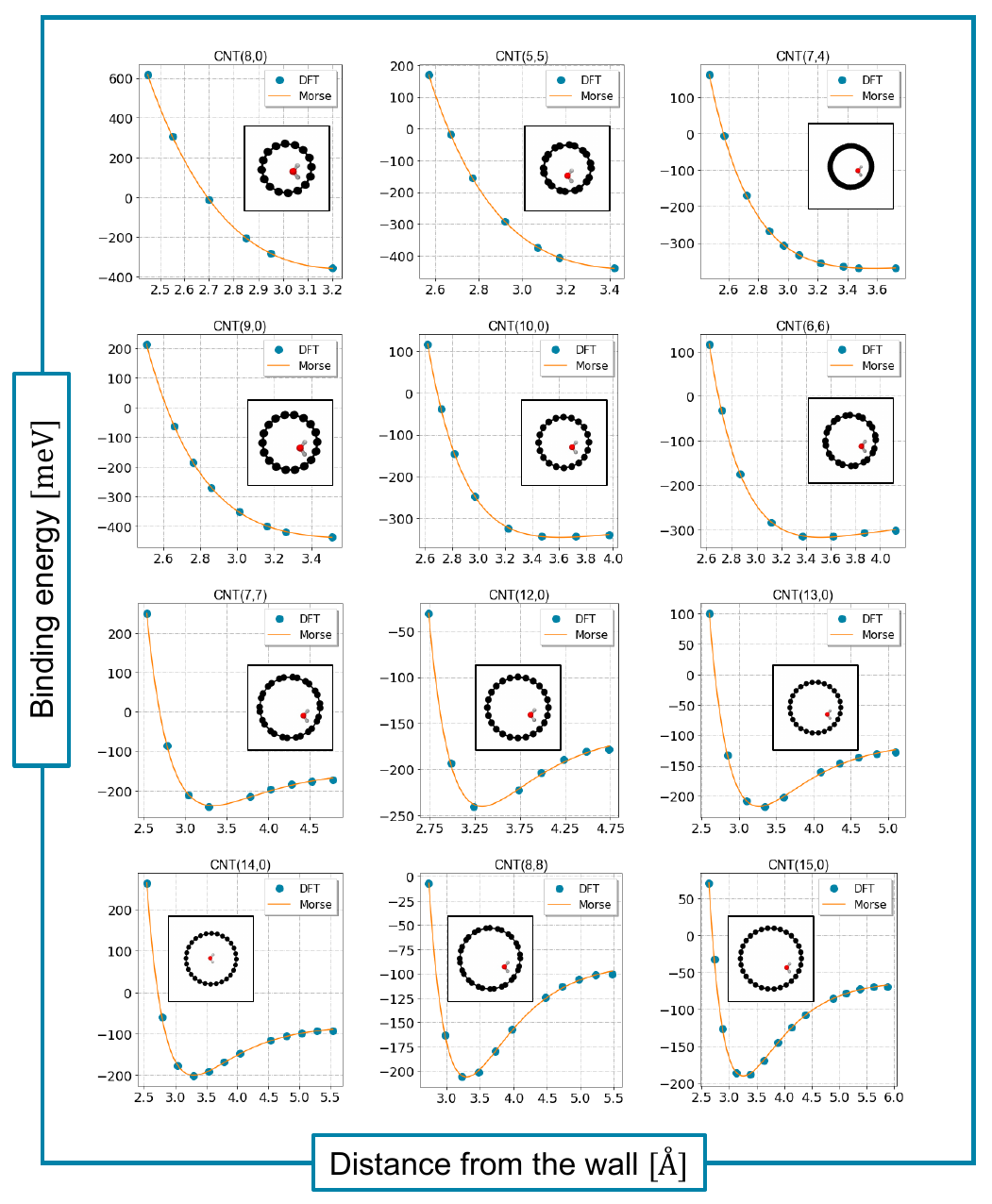}
    \caption{Binding energy of a single water molecule inside a CNT as function of the distance from the CNT wall. The DFT points are computed as the average of the DFT binding energy of the water molecule in the 12 considered directions. The parameters of the fitted Morse potentials are reported in Table \ref{si:tab-morse}. The insets show a snapshot of the water molecule inside the CNT.}
    \label{fig:si-confining-potential}
\end{figure}
\noindent The confining potentials are written as 
\begin{equation}
    U_{\text{Morse}}(d_w) = E_0 + D_e \left(1 - e^{-a(d_w-R_e)}\right)^2,
    \label{si:eq-morse}
\end{equation}
where $d_w$ is the distance from the wall, and $E_0$ is the binding energy at the equilibrium distance $R_e$. Note that $D_e$ and $E_0$ do not coincide because of the constraint on the distance from the wall $d_w$, which cannot be greater than the nanotube radius.  
The parameters of the confining Morse potential fitted to the DFT data for each CNT are reported in Table \ref{si:tab-morse}. Using the computed DFT data, we subsequently extrapolated the Morse confining potential for a larger nanotube, CNT(16,0), with a linear fitting based on the parameters of CNT(14,0), CNT(8,8) and CNT(15,0). 
The parameters of the Morse potential for CNT(16,0) are reported in the last row of Table \ref{si:tab-morse}.

\begin{table}[tbh!]
    \centering
    \begin{tabular}{|l|l|l|l|l|}
    \hline
        ~ & $D_e$ (meV) & $a$ ($\text{\AA}^{-1}$) & $R_e$ ($\text{\AA}$)& $E_0$ (meV) \\ \hline 
        (8,0) & 778.9187456 & 0.961092979 & 3.231366026 & -357.8447059 \\ \hline
        (5,5) & 148.3094582 & 1.186533705 & 3.504812408 & -441.945131 \\ \hline
        (7,4) & 29.33497992 & 1.507964732 & 3.571405126 & -369.8974892 \\ \hline
        (9,0) & 39.35735409 & 1.428163409 & 3.646583554 & -439.3356747 \\ \hline
        (10,0) & 33.66605706 & 1.574361596 & 3.602931175 & -345.0447328 \\ \hline
        (6,6) & 47.20301471 & 1.551865302 & 3.518225692 & -317.2421718 \\ \hline
        (7,7) & 89.29679021 & 1.506233119 & 3.330121661 & -237.9370359 \\ \hline
        (12,0) & 88.41756996 & 1.490191514 & 3.361055153 & -240.0209542 \\ \hline
        (13,0) & 106.3736042 & 1.496821862 & 3.270211774 & -216.7653559 \\ \hline
        (14,0) & 120.6571729 & 1.431669518 & 3.29813939 & -200.1075625 \\ \hline
        (8,8) & 118.9349315 & 1.42934899 & 3.306555169 & -204.8066963 \\ \hline
        (15,0) & 129.7962591 & 1.386600161 & 3.263996583 & -189.5961614 \\ \hline
        (16,0) & 129.3701381 & 1.413343892 & 3.255360094 & -190.4070647 \\ \hline
    \end{tabular}
    \caption{Parameters of the Morse potential in Eq.\ \ref{si:eq-morse} fitted to the DFT binding energy of a water molecule (in meV) inside a CNT as a function of the distance from the CNT wall (in $\text{\AA}$). The first column reports the chirality of the CNT(n,m). The DFT energies and the fitted confining potentials are plotted in Fig.\ \ref{fig:si-confining-potential}. The Morse potential for the largest nanotube (n=16) are extrapolated with a linear fitting on the parameters for CNT(14,0), CNT(8,8) and CNT(15,0).}
    \label{si:tab-morse}
\end{table}

\clearpage

\begin{figure}[tbh!]
    \centering
    \includegraphics[scale=0.4]{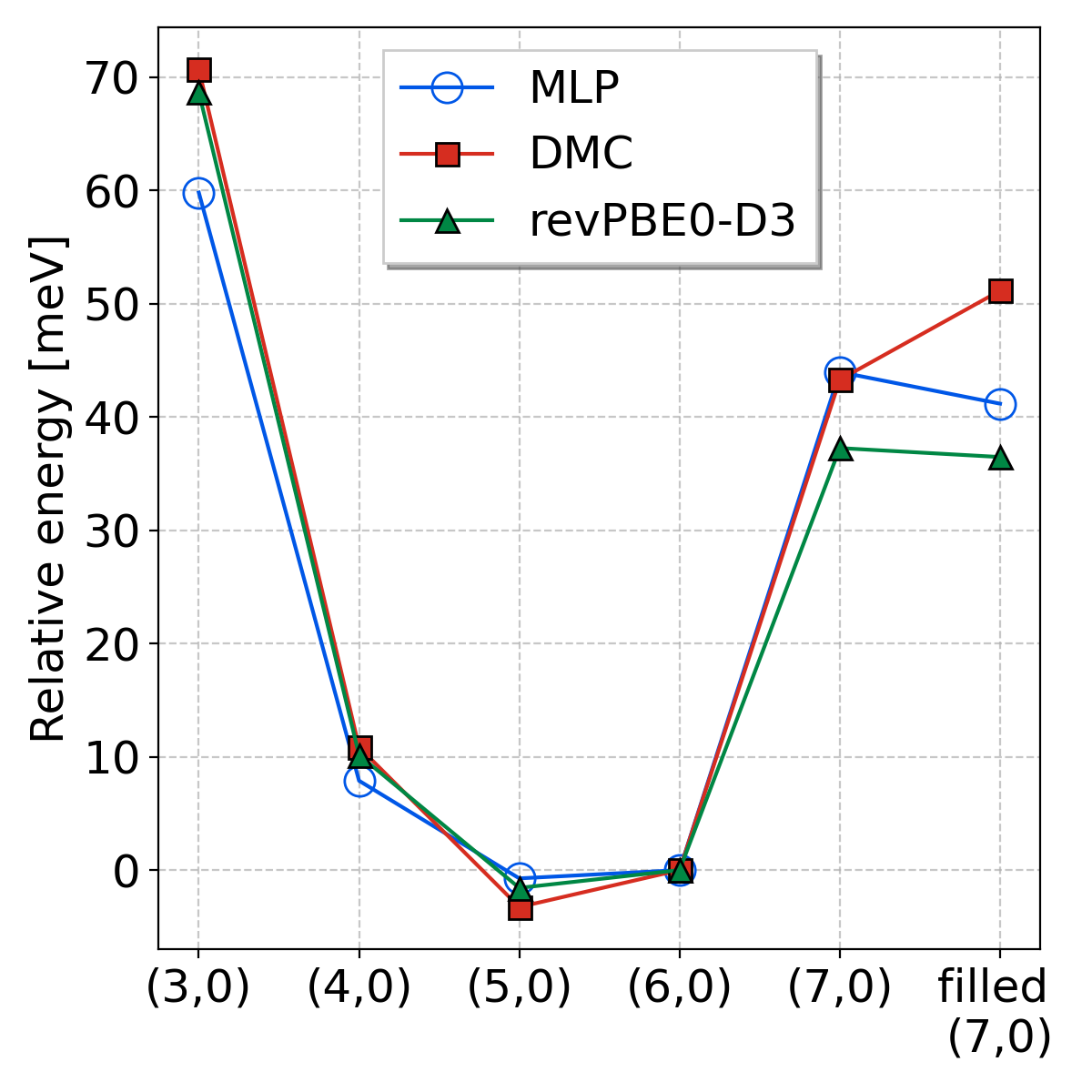}
    \caption{Validation of the MLP in the RSS. The figures shows the relative energies of the metastable phases found with the RSS algorithm for $d\sim 11.8\text{ \AA}$ with: the MLP (blue); DMC (red); and the reference DFT functional revPBE0-D3 (green). The MLP captures the relative energies with a maximum error of $\sim 10 \text{ \AA}$.}
    \label{fig:si_RSS}
\end{figure}

\numberedsection{Validation of the MLP with DMC}\label{si-sec:dmc}
The  metastable ice polymorphs in each confining nanotube are determined by combining the MLP model with a RSS approach. The structure stable at zero pressure and zero temperature are identified by the minimum in the relative energy. To validate our model, we compare the MLP relative energies for the water-water interaction to DMC data (Fig.\ \ref{fig:si_RSS}) for the diameter $d \sim 11.8 \text{ \AA}$. The DFT data with revPBE0-D3 (on which the MLP is trained) are also shown. In particular, for each phase we report the relative energy, i.e. the difference between the energy of each phase and ice (6,0).
Overall, the difference in the energetics prediction of the metastable phases with the MLP and DMC are less than $\sim 10 \text{ meV}$.

\begin{figure}[tbh!]
    \centering
    \includegraphics[scale=0.5]{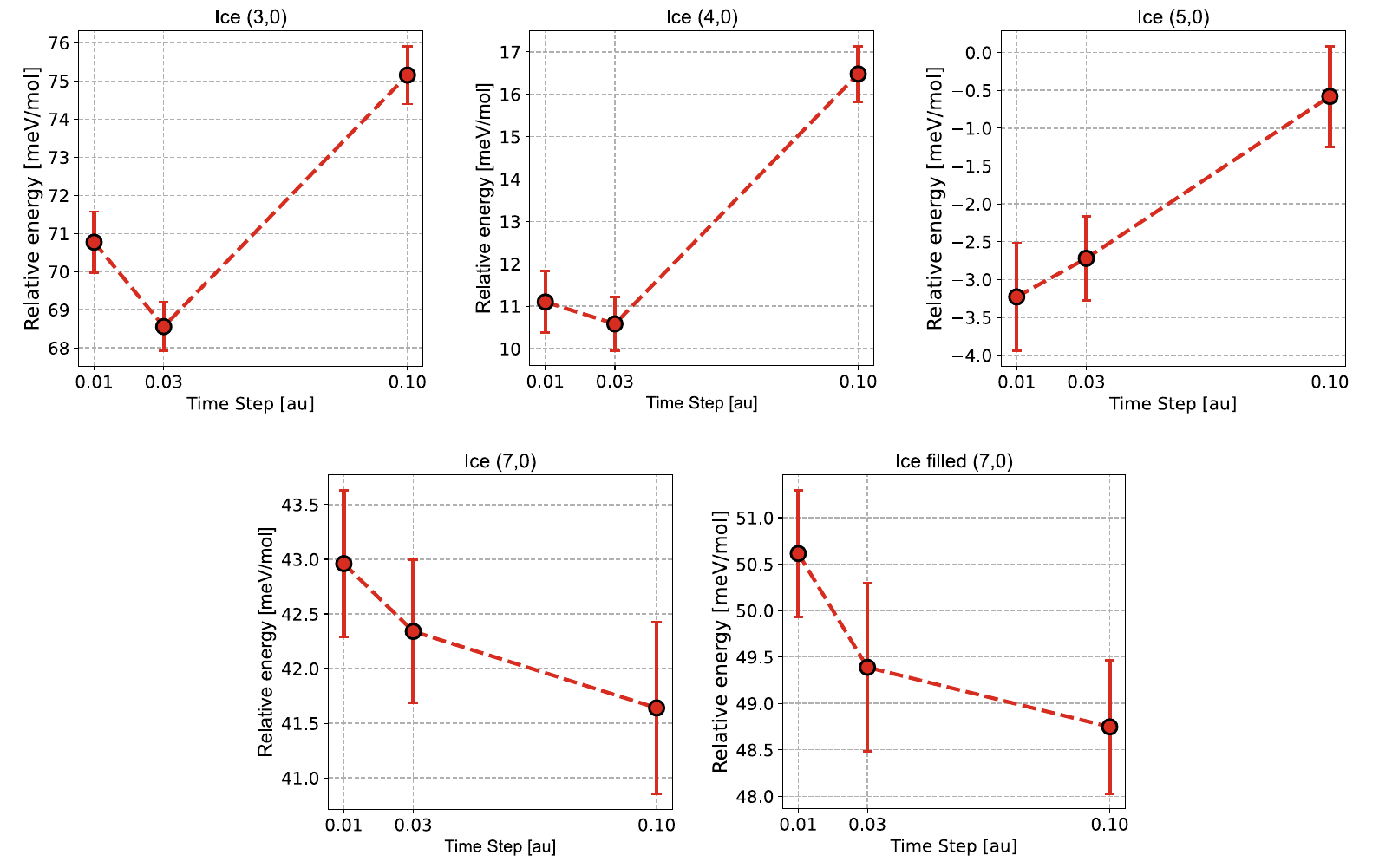}
    \caption{Convergence of the DMC relative energies with respect to the time step. Each plot reports the difference between the energy of the specific phase and ice (6,0) as a function of the time step. The error bars are the statistical error bars of the DMC simulations.}
    \label{fig:si_dmc_time_step}
\end{figure}

The time step $\tau$ is a key factor affecting the accuracy of DMC calculations. In DMC, a propagation according to the imaginary time Schrödinger equation is performed to project out the exact ground state from a trial wave function \cite{QuantumMonteCarlo}. A time step $\tau$ must be chosen, but the projection is exact only in the continuous limit $\tau \to 0$. However, the ZSGMA \cite{ZSGMA} DMC algorithm gives better convergence with respect to $\tau$ than previously used methods. In this work, we have verified the time step convergence for each analyzed system. The time step convergence of the relative energies reported in Fig.\ \ref{fig:si_RSS} are plotted in Fig.\ \ref{fig:si_dmc_time_step}. The values used in Fig.\ \ref{fig:si_RSS} are computed with a time step of $0.01 \text{ a.u}$.

\clearpage
\numberedsection{Convergence of coexistence simulations}\label{si-sec:coexistence-simulations}

The melting temperature of the ice nanotubes is determined via solid-liquid coexistence simulations. The initial interface is built by melting half of the ice nanotubes at high temperatures in the NVT ensemble, as described in the \textbf{Methods} section of the main manuscript. Subsequently we run coexistence simulations in the NP$_z$T ensemble changing the temperature at the fixed pressure $P\sim 10 \text{ bar}$. The simulated cells contain respectively $2130$ atoms for $d\sim 9.5 \text{ AA}$, $2160$ atoms for $d\sim 10.2 \text{ \AA}$, $2100$ atoms for $d\sim 11.0 \text{ \AA}$, $2286$ atoms for $d\sim 11.8 \text{ \AA}$, and $2304$ atoms for $d\sim 12.5 \text{ \AA}$. An overview of the systems investigated in this work is reported in Table \ref{si:tab-setup}. Tests on the convergence of the density with respect to the system size are reported in section \ref{si-sec:fse}.

\begin{table}[h!]
\centering
\begin{tabular}{|p{4cm}|c|}
\hline
\multicolumn{1}{|c|}{Simulation details} & Illustration \\ \hline \hline
\centering d $\sim$ 9.5 $ \text{ \AA}$\newline N$_{\text{water}}$ = 2130 & \raisebox{-0.5\height}{\includegraphics[width=6cm]{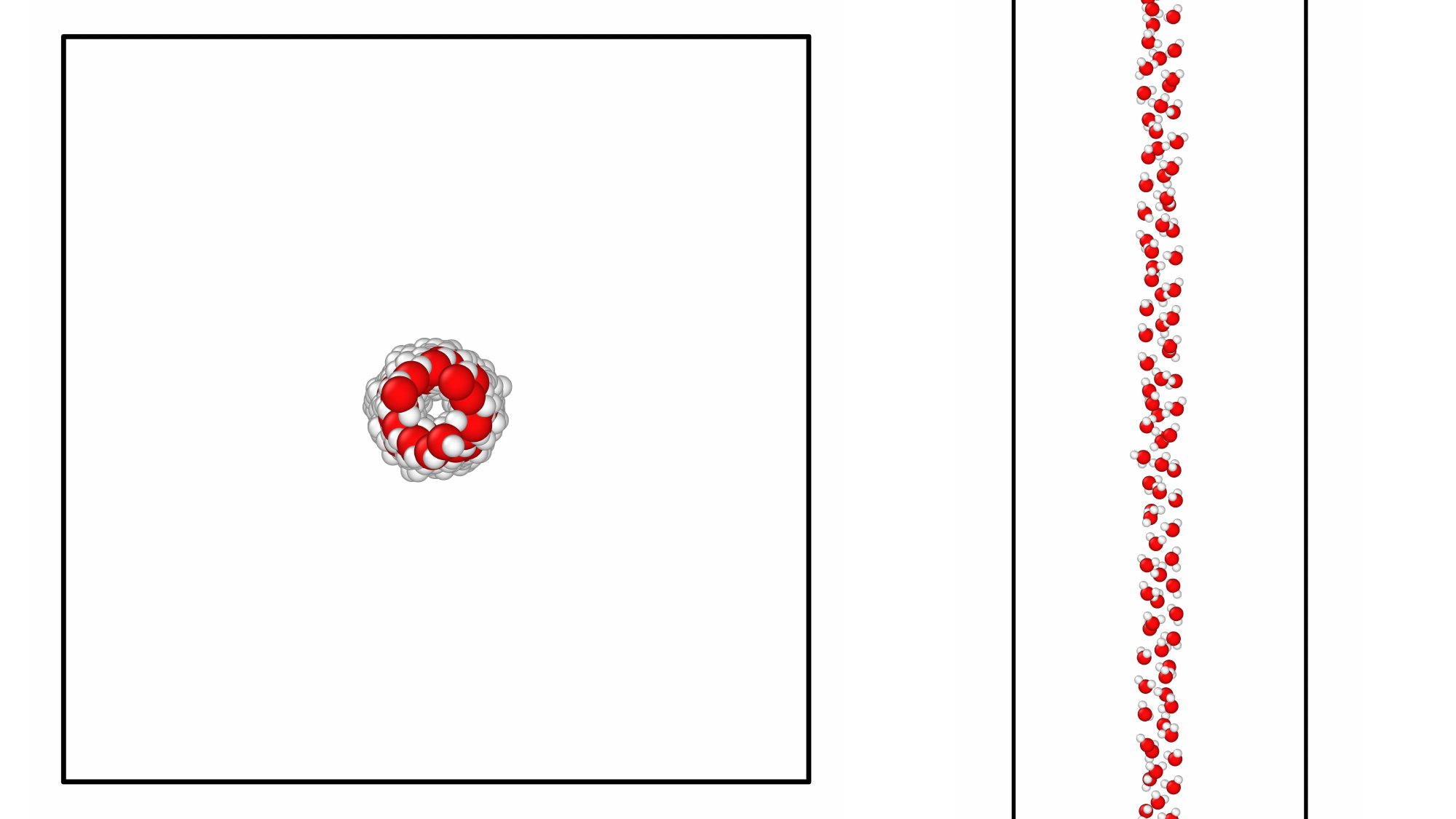}} \\ \hline \hline
\centering d $\sim$ 10.2 $ \text{ \AA}$\newline N$_{\text{water}}$ = 2160 & \raisebox{-0.5\height}{\includegraphics[width=6cm]{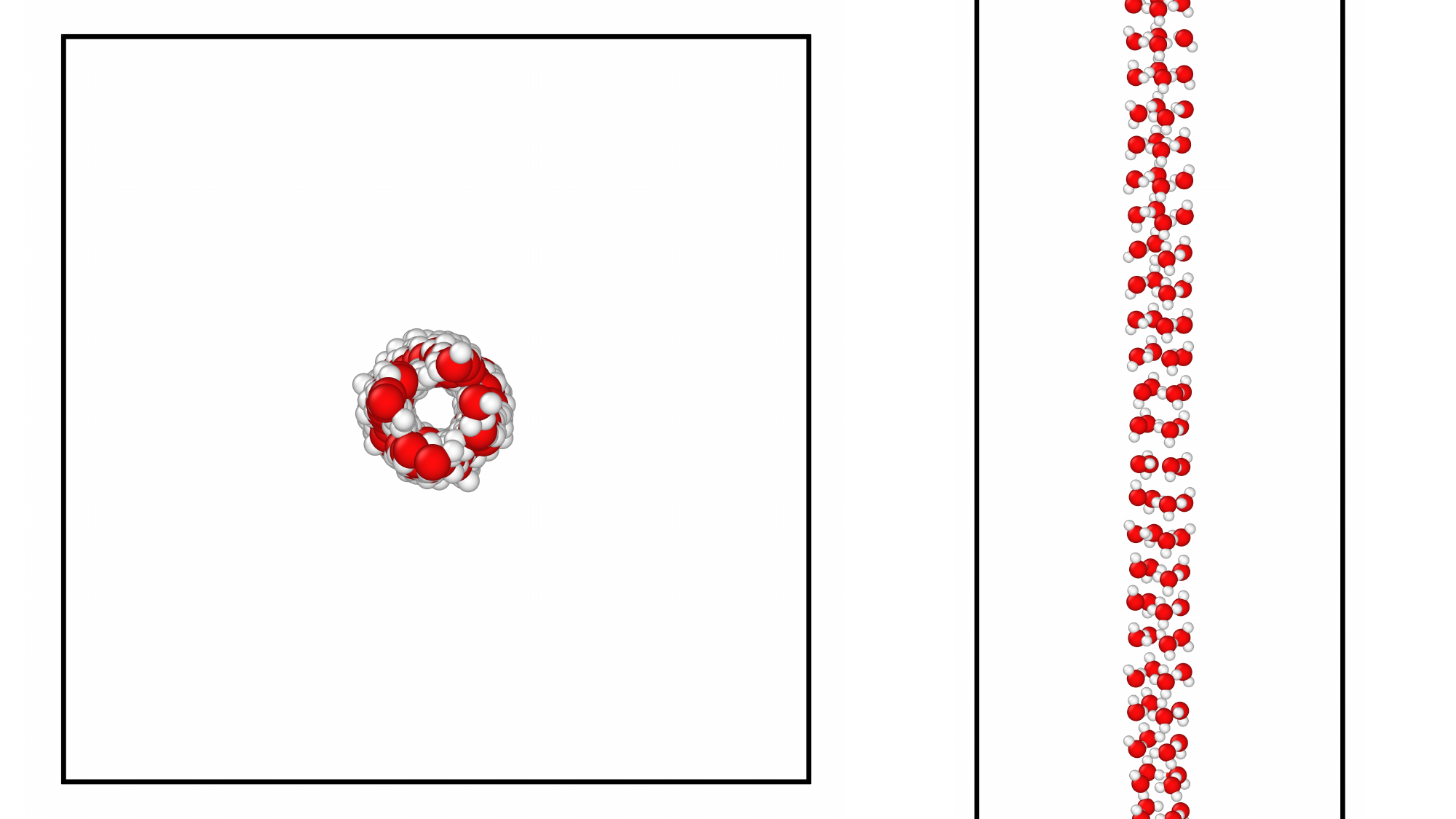}} \\ \hline \hline
\centering d $\sim$ 11 $ \text{ \AA}$\newline N$_{\text{water}}$ = 2100 & \raisebox{-0.5\height}{\includegraphics[width=6cm]{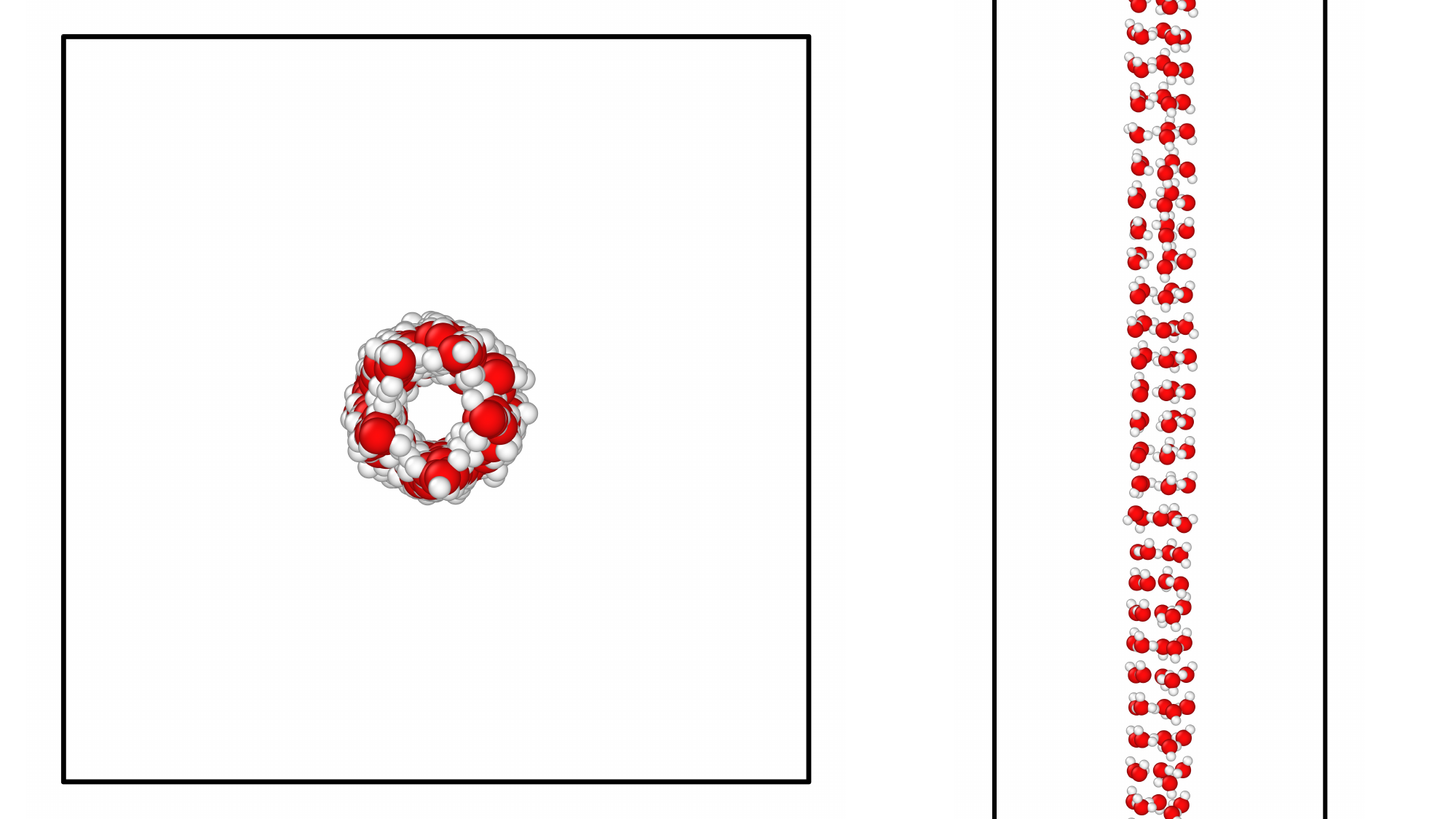}} \\ \hline \hline
\centering d $\sim$ 11.8 $ \text{ \AA}$\newline N$_{\text{water}}$ = 2286 & \raisebox{-0.5\height}{\includegraphics[width=6cm]{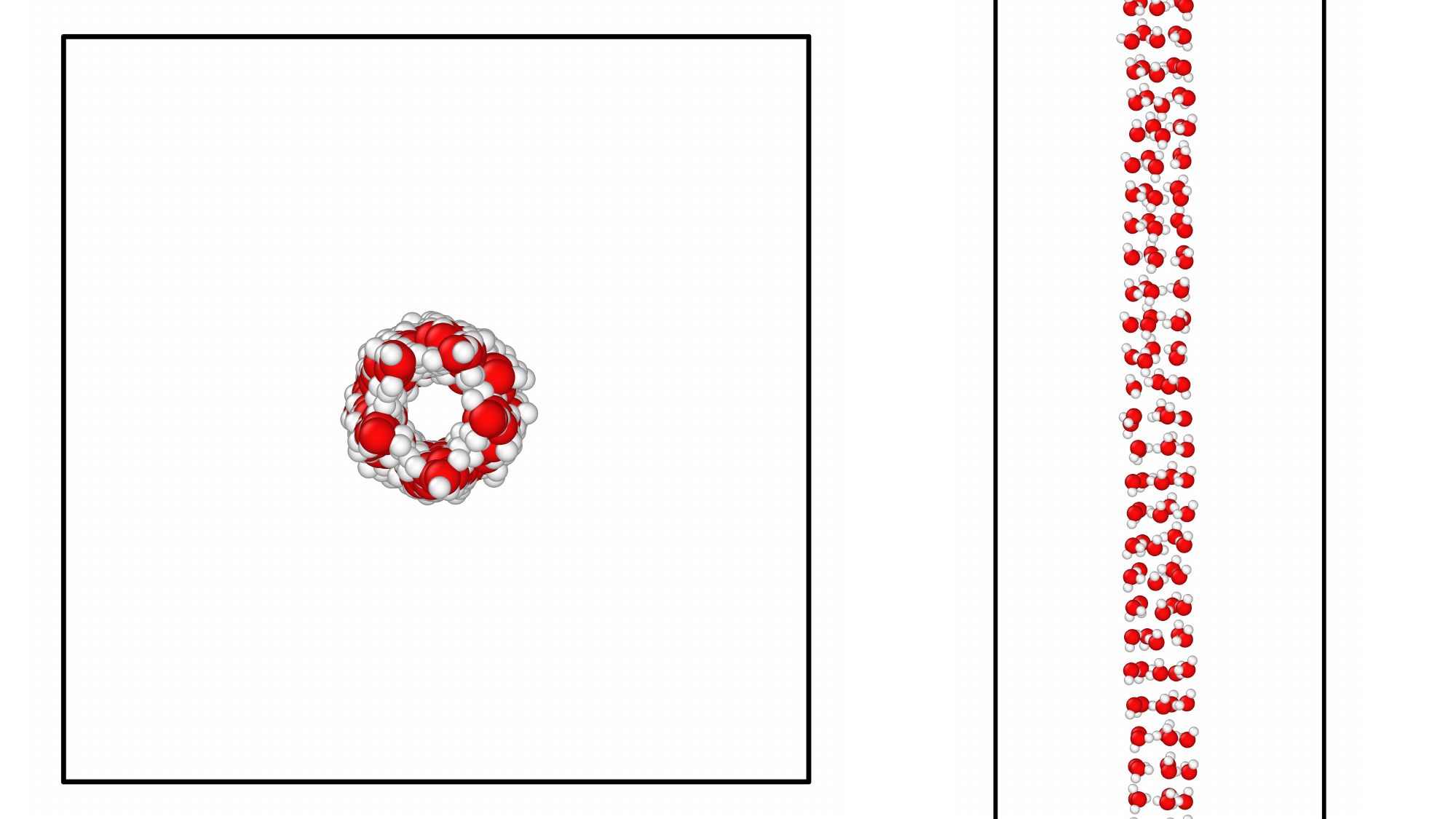}} \\ \hline \hline
\centering d $\sim$ 12.5 $ \text{ \AA}$\newline N$_{\text{water}}$ = 2304 & \raisebox{-0.5\height}{\includegraphics[width=6cm]{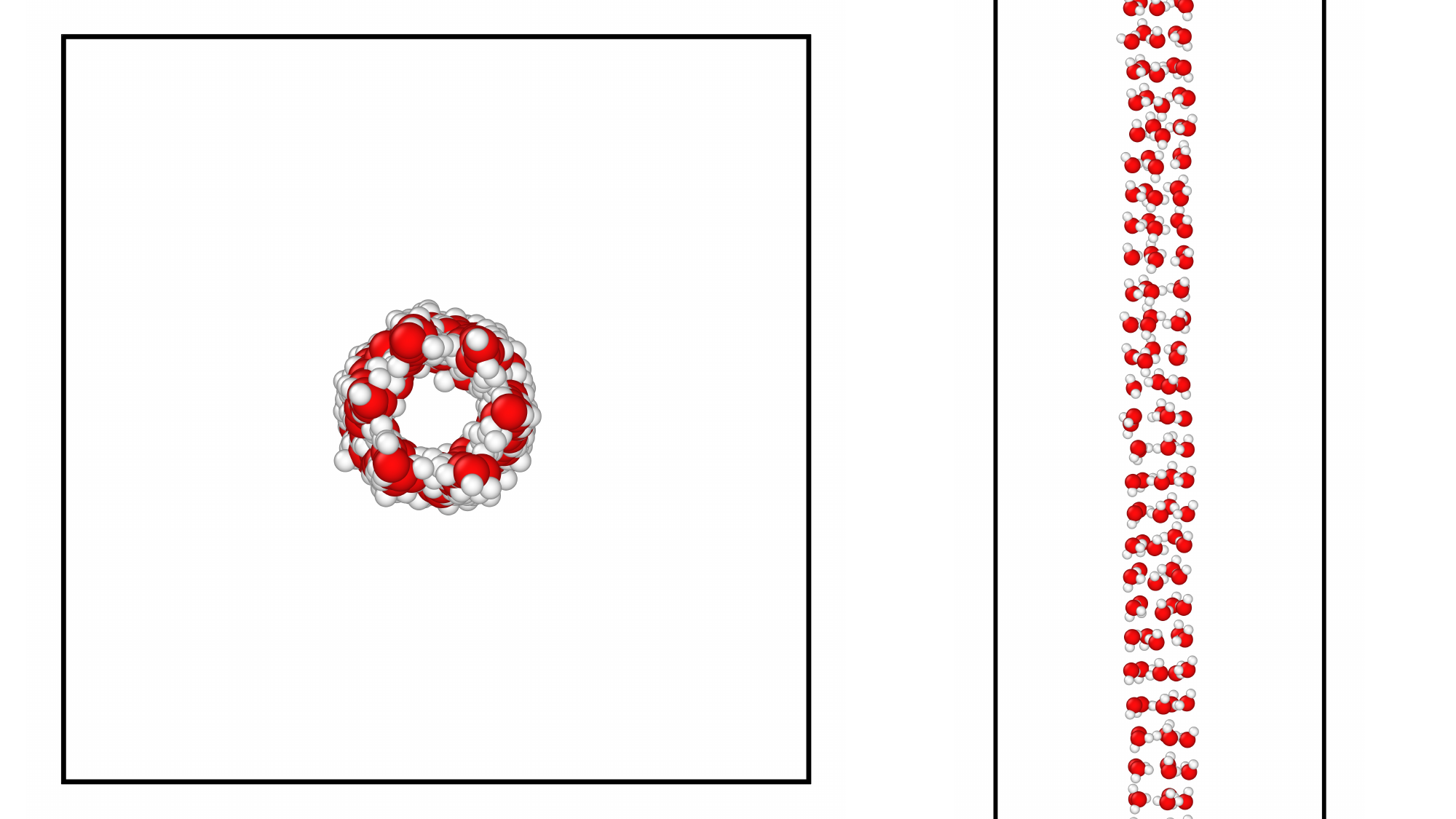}} \\ \hline
\end{tabular}
\caption{Overview of the systems investigated in this work. For each system, we report the confinining diameter $d$, and the total number of water molecules N$_\text{water}$ in the coexistence simulations. In the right column, we show a front and side view of each system at the temperature $T = 250 \text{ K}$. The black lines represent the edges of the simulation box.}
    \label{si:tab-setup}
\end{table}

\begin{figure}[tbh!]
    \centering
    \includegraphics[scale=0.3]{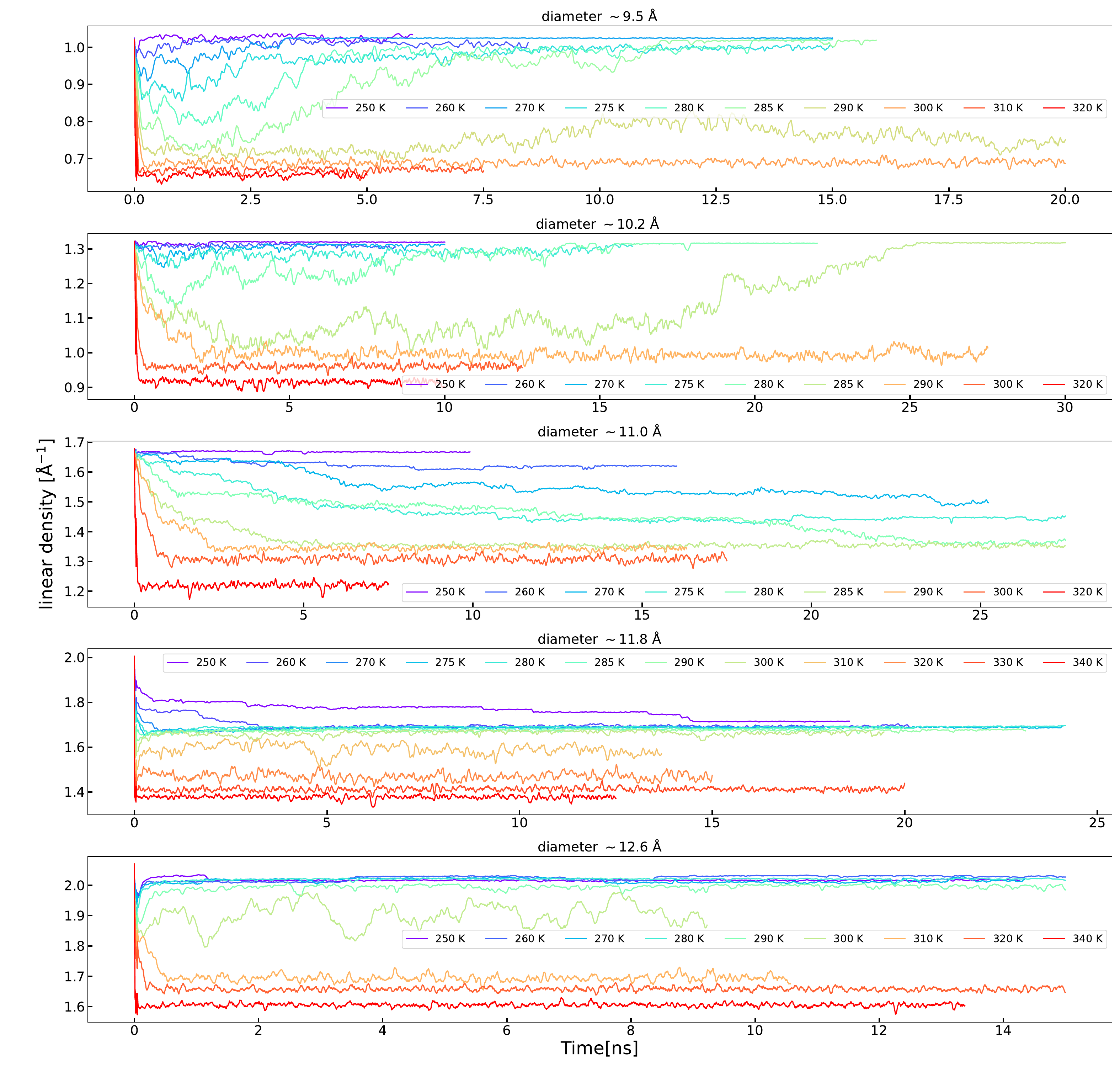}
    \caption{Coexistence simulations. Running averages of the linear density (number of molecules divided by the length of the simulation box) as a function of time at each temperature and for each confining nanotube.}
    \label{fig:si-convergence-density}
\end{figure}

In Fig.\ \ref{fig:si-convergence-density}, we plot the linear density (number of molecules divided by the length of the cell) as a function of the simulation time. Relatively short simulations ($\sim 6 \text{ ns}$) are necessary to equilibrate the solid at low temperatures and the liquid at high temperatures. Longer simulations ($> 20 \text{ ns}$) are necessary to achieve convergence for $T$ close to the melting temperature.

\clearpage
\numberedsection{Estimate of the diffusion coefficient}\label{si-sec:diffusion-coefficient}
In this section, we report an analysis on the estimate of the diffusion coefficient of the ice nanotubes as a function of the temperature. 
The two most common methods used to estimate the diffusion coefficient in molecular dynamics (MD) simulations are: (1) using the Einstein relation to extract the diffusion coefficient from the Mean Square Displacement (MSD)\cite{frenkel2023understanding}; (2) using the Green-Kubo relation to extract the diffusion coefficient from the Velocity Autocorrelation Function (VACF)\cite{frenkel2023understanding}.

In the first method, the diffusion coefficient is related to the MSD of a particle as a function of the observation time. In particular, the diffusion coefficient is proportional to the observation time in the limit that the observation time goes to infinity:
\begin{equation}\label{si:eq-msd}
    D = \frac{1}{2n_d} \lim_{t \to \infty} \frac{\left\langle \left[ \mathbf{r}(t_o+t)-\mathbf{r}(t_o) \right]^2\right\rangle}{t},
\end{equation}
where $D$ is the diffusion coefficient, $\mathbf{r}(t)$ is the particle position at time $t$, and $n_d$ is the dimensionality of the system. The numerator in equation \ref{si:eq-msd} is the ensemble average of the particle square displacement. The ensemble average is an average over all the particles in the simulation and all the time origins $t_o$. The diffusion coefficient $D$ can easily be obtained from the slope of the curve MSD versus time divided by $2d$.

The second method is based on linear response theory and estimates the diffusion coefficient from equilibrium MD simulations using a Green-Kubo relation:
\begin{equation}\label{si:eq-vacf}
    D = \frac{1}{n_d} \int_0^t \left\langle \mathbf{v}(t')\cdot \mathbf{v}(0) \right\rangle dt',
\end{equation}
where $\mathbf{v}(t)$ is the velocity at time $t$, and the brackets correspond to an ensemble average.

\begin{figure}[tbh!]
    \centering
    \includegraphics[scale=0.5]{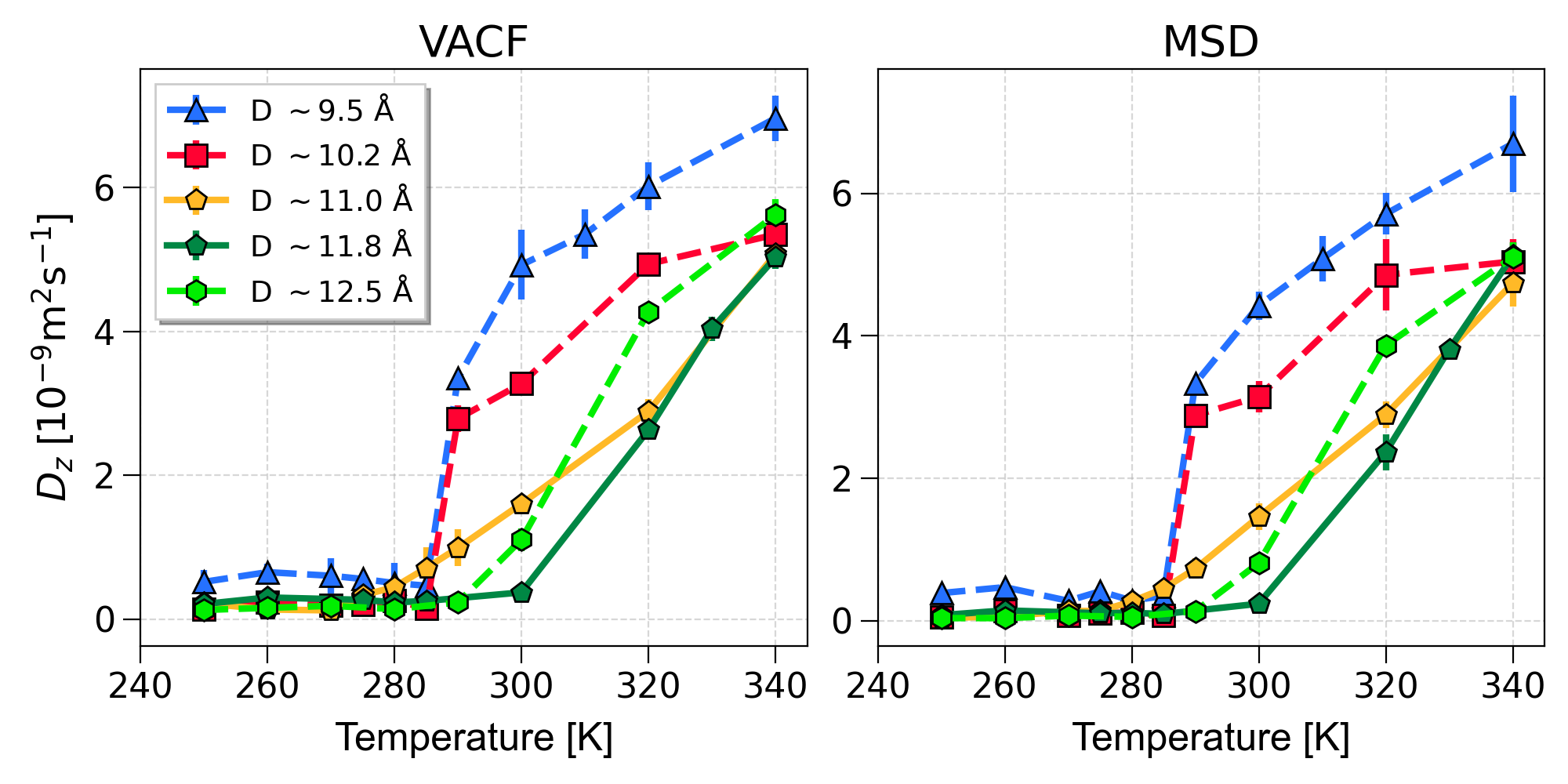}
    \caption{Diffusion coefficient as a function of the temperature with the VACF (left) and the MSD (right) method. The error bars due to the statistical sampling are estimated via bootstrapping.}
    \label{fig:si-diffusion_msd_vs_vacf}
\end{figure}

\begin{figure}[tbh!]
    \centering
    \includegraphics[scale=0.5]{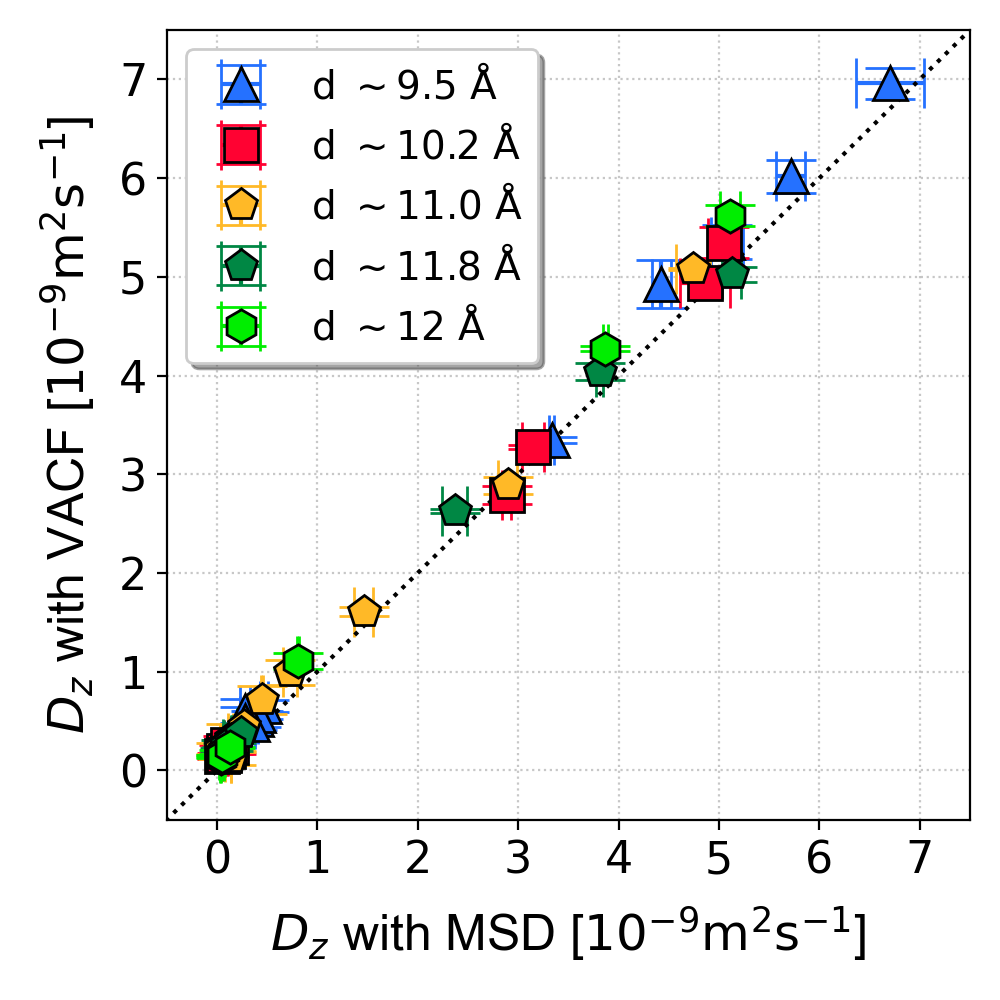}
    \caption{Scatter plot of the diffusion coefficient computed with the VACF ($y$ axis) and the MSD ($x$ axis) method. The error bars due to the statistical sampling are estimated via bootstrapping.}
    \label{fig:si-scatter_diffusion_msd_vs_vacf}
\end{figure}

In this work, we estimated the oxygen diffusion coefficient along the nanotube axis $D_z$ as a function of the temperature for each confining diameter applying both methods to MD trajectories in the NVT ensemble (i.e., constant number of particles N, volume of the box V and temperature T). 
The MD-NVT trajectories are computed with a $0.5 \text{ fs}$ time step and the `gle' thermostat, starting from the final configuration of the equilibrated coexistence simulations in the NP$_z$T ensemble. The MSD estimates are obtained with $0.5 \text{ ns}$ long trajectories, computing the MSD with a $20 \text{ ps}$ correlation time. The slope of the MSD is estimated with a linear fitting between $\sim 5 \text{ ps}$ and $\sim 15 \text{ ps}$.
The VACF estimates are obtained with $160 \text{ ps}$ long trajectories for the solid phase and $320 \text{ ps}$ long trajectories for the liquid phase. The VACF is computed with a $10 \text{ ps}$ correlation time.
As shown in Fig.\ \ref{fig:si-diffusion_msd_vs_vacf} and \ref{fig:si-scatter_diffusion_msd_vs_vacf}, the two methods are consistent within the statistical error bar.

The error bars on the diffusion coefficient due to the statistical sampling were computed as 2 standard deviations, which are estimated via bootstrapping by dividing the trajectory in two blocks.

\numberedsection{Implicit vs explicit carbon}\label{si-sec:explicit}
In the main manuscript, we focus on the melting temperature of quasi one-dimensional ice polymorphs confined in a uniform cylindrical potential. The confining potential is fitted to the DFT water-carbon interaction (as described in Sec.\ \ref{si-sec:confining-potential}) inside a CNT. We refer to this model as an \textit{implicit} carbon model.

In this section, we analyse the effect of \textit{explicit} carbon on the results reported in the main manuscript. 
In particular, we considered the MLP trained on DFT revPBE-D3 data for water inside CNTs from Ref.\ \citenum{Thiemann_Schran_water_flow}. With this potential, we computed the diffusion coefficient as a function of the temperature in CNT(14,0), and both the diffusion coefficient and the number of hydrogen bonds at $T=320 \text{ K}$ for CNT(14,0), CNT(15,0), and CNT(16,0).
The simulations with the explicit model were run as follows: (1) we started from an initial configuration resembling the water density of the implicit model in the NP$_z$T simulation; (2) we run a $\sim 2 \text{ ns}$ NVT simulation at the same temperature to estimate the number of hydrogen bonds and the diffusion coefficient. The length of the CNTs used in the explicit model simulations is $\sim 240 \text{ \AA}$, corresponding to a number of water molecules of $\sim 300$.

\begin{figure}[tbh!]
    \centering
    \includegraphics[scale=0.45]{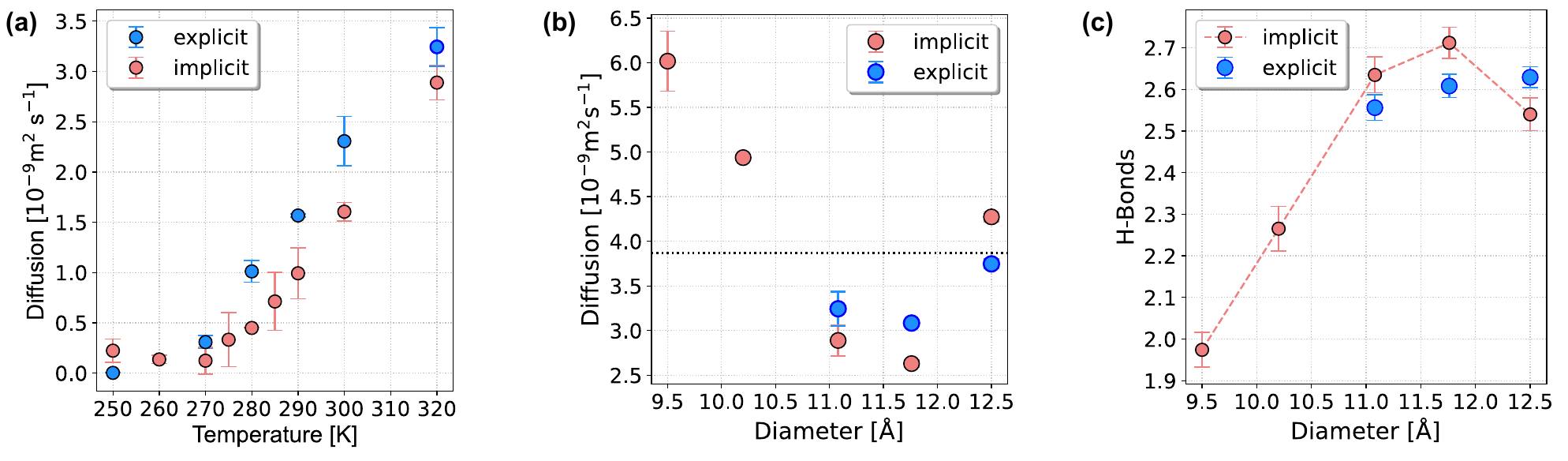}
    \caption{Implicit vs explicit carbon model. (a) Diffusion coefficient as a function of temperature in CNT(14,0). (b) Diffusion coefficient as a function of the diameter at $320 \text{ K}$. (c) Number of hydrogen bonds as a function of the diameter at $320 \text{ K}$. The error bars due to the statistical sampling are estimated via bootstrapping.}
    \label{fig:si-explicit-model}
\end{figure}

In Fig.\ \ref{fig:si-explicit-model} (a) we report the diffusion coefficient as a function of the temperature for the CNT(14,0) with the implicit (red) and the explicit (blue) model. With the explicit model, we observe a slightly enhanced diffusion at fixed temperature in the liquid phase compared to the implicit model. 
The enhanced diffusion can be physically ascribed to the coupling with the phonon modes of the carbon, as shown in Refs. XXX. Furthermore, little differences between the two models are expected due to the use of different DFT functionals in the training of the MLPs. 
Panels (b) and (c) of Fig.\ \ref{fig:si-explicit-model} show respectively the diffusion coefficient and the number of hydrogen bonds of the liquid phase ($T=320 \text{ K}$) as a function of the diameter for the three largest nanotubes considered in the main manuscript. As in the previous case, we observe little differences between the explicit and implicit model.
Overall, this test show that the implicit model is fairly accurate in describing the dynamics of the ice nanotubes. The difference in the prediction of the melting temperature with the explicit carbon model is $\sim 15 \text{ K}$.

\begin{figure}[tbh!]
    \centering
    \includegraphics[scale=0.5]{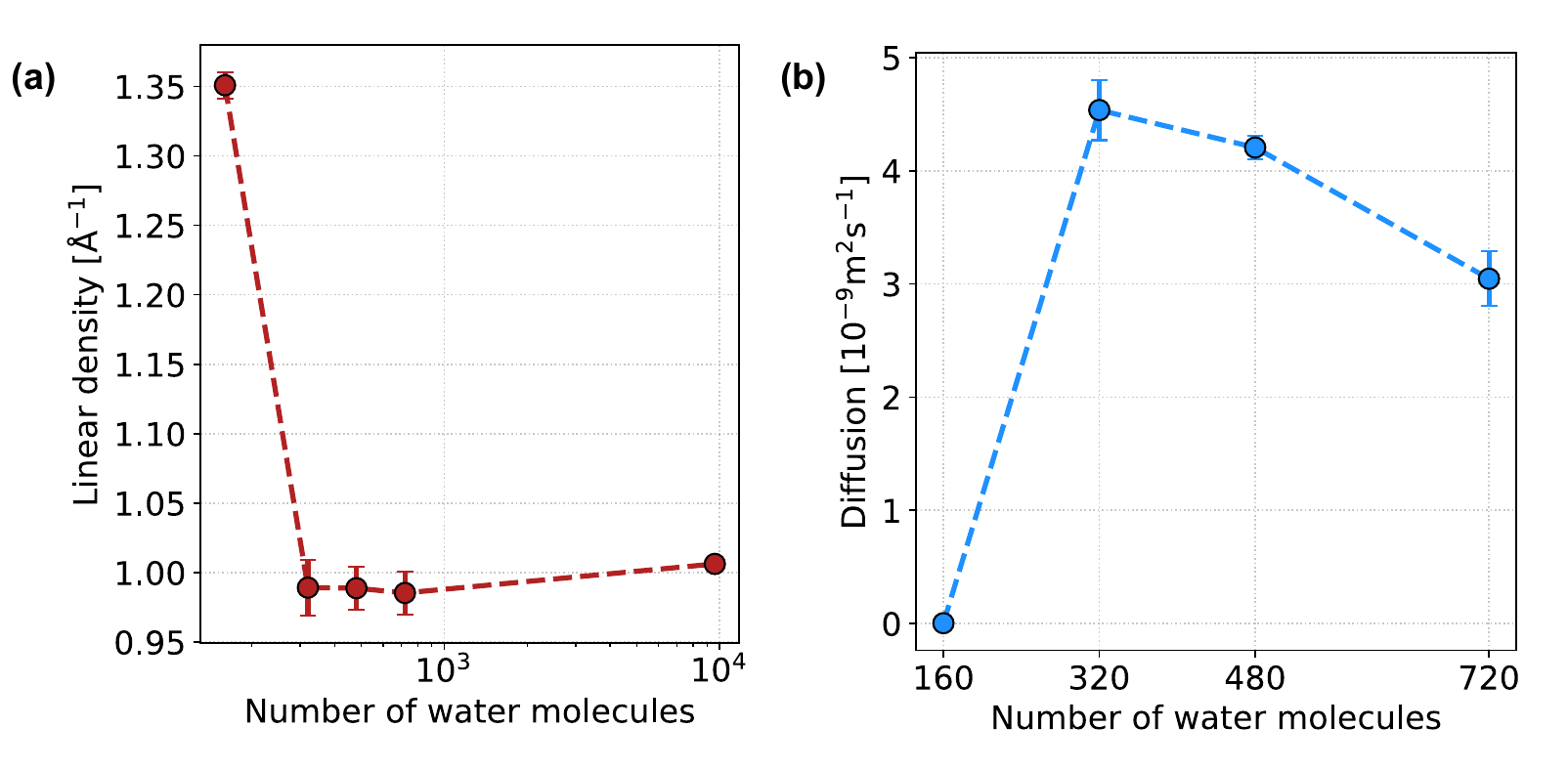}
    \caption{Test on finite size errors for $d \sim 10.2 \text{ \AA}$ and $T=300 \text{ K}$. (a) Linear density (number of water molecules divided by the length of the simulation box) as a function of the number of water molecules. (b) Diffusion coefficient along the nanotube axis as a function of the number of water molecules. The error bars due to the statistical sampling are estimated via bootstrapping.}
    \label{fig:si-fse-1}
\end{figure}

\numberedsection{Finite size errors}\label{si-sec:fse}
In this section, we report a test on the finite size errors on the estimate of the density and the diffusion coefficient, used in the main manuscript to characterize the phase transition and identify the melting temperature. In particular, we consider the case of square ice (diameter $d \sim 10.2 \text{ \AA}$). 
In Fig.\ \ref{fig:si-fse-1}, we plot the linear density (a) and the diffusion coefficient (b) as a function of the number of water molecules $N_\text{mol}$ at the fixed temperature $T=300 \text{ K}$. In particular, we consider $N_\text{mol} = 160, 320, 480, 720,$ and $9600$ (except for the diffusion coefficient, where longer simulations are necessary to achieve convergence). Noticeably, the largest number of water molecules simulated ($N=9600$), is comparable to the number of water molecules expected in a realistic experimental set-up\cite{Exp_photoluminescence_Homma} with a CNT of length $~ 1 \mu\text{m}$ (assuming the same linear density of water molecules of $\sim 1 \text{ \AA}^{-1}$).

\begin{figure}[tbh!]
    \centering
    \includegraphics[scale=0.5]{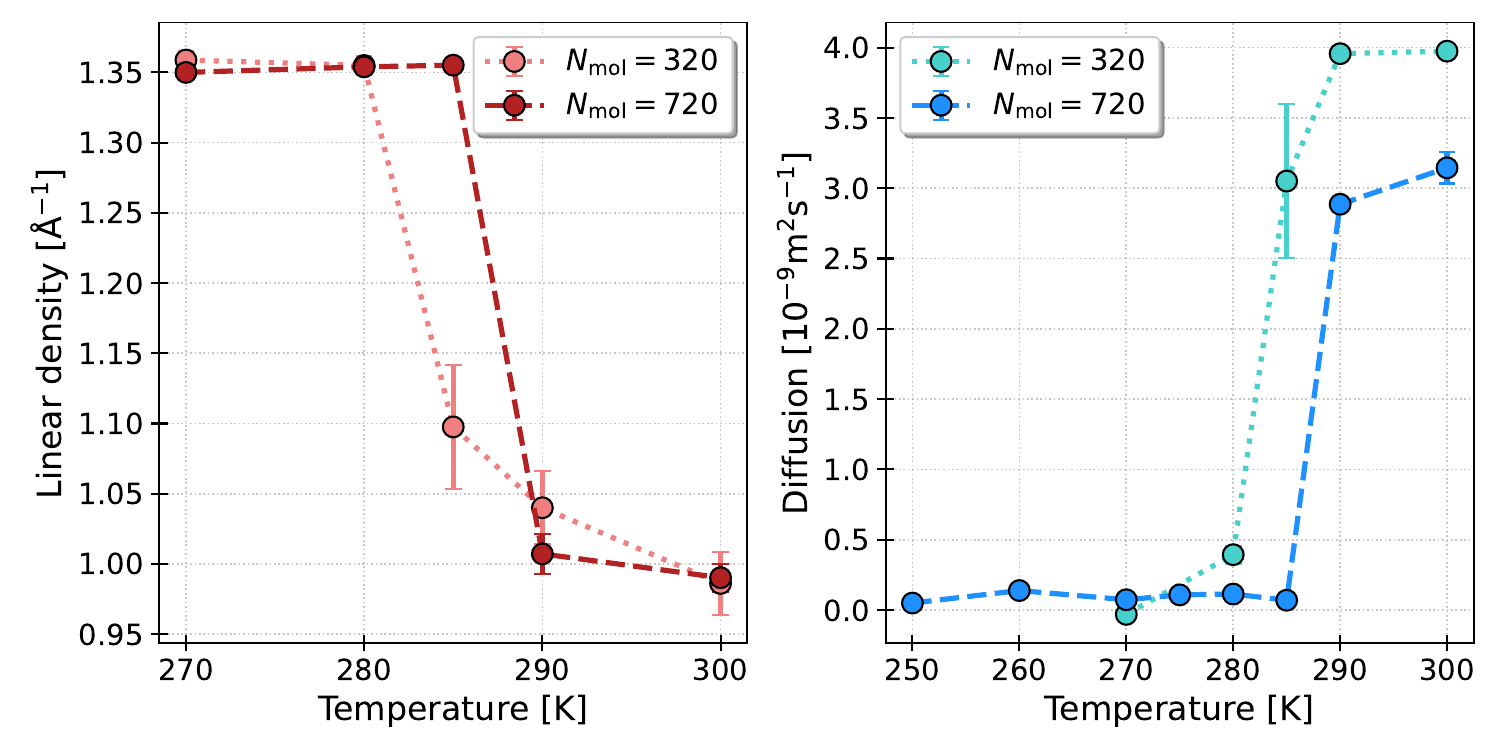}
    \caption{Test on finite size errors on the melting temperature for $d \sim 10.2 \text{ \AA}$. Linear density (left) and diffusion coefficient (right) as a function of the temperature for $N_\text{mol}=320$. The melting temperature predicted with $320$ water molecules is $T \sim 280 - 285\text{ K}$; the melting temperature reported in the main manuscript considering $720$ water molecules is $T \sim 285-290 \text{ K}$. The error bars due to the statistical sampling are estimated via bootstrapping.}
    \label{fig:si-fse-2}
\end{figure}

In Fig.\ \ref{fig:si-fse-2} we plot the density (left) and the diffusion coefficient (right) as a function of the temperature for $320$ and $720$ water molecules. The difference in the melting temperature predicted with $320$ and $720$ water molecules is $\sim 5 \text{ K}$.

Overall, our tests show that the simulations  $\sim 10^3$ water molecules considered in the main manuscript are converged for the computational of the melting temperature and the structural/dynamical properties analysed in this work.

\begin{figure}[tbh!]
    \centering
    \includegraphics[scale=0.5]{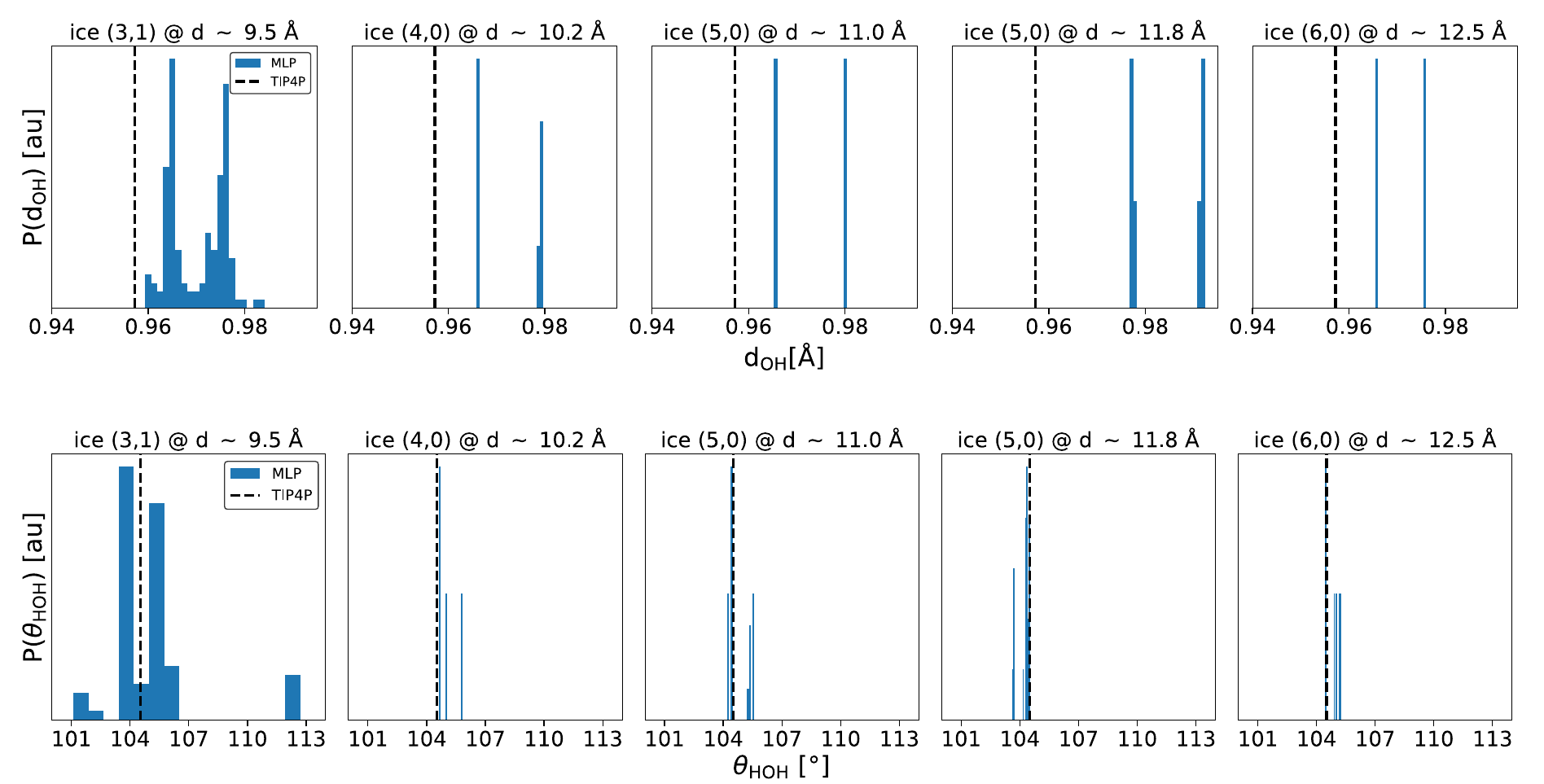}
    \caption{Histograms of O-H distances and H-O-H angles for the optimized structures of triangular, square, pentagonal and hexagonal ice. The blue bars represent the result with our MLP. The black dashed vertical line is the corresponding fixed value for the TIP4P original rigid model.}
    \label{fig:si-distances_angles}
\end{figure}

\clearpage
\numberedsection{Distributions of bonds and angles}\label{si-sec:distances_angles}
In this section, we report the distributions of the O-H distances and the H-O-H angles in the MLP optimized structures considered in our manuscript. The histograms of the distribution are plotted in Fig.\ \ref{fig:si-distances_angles}. For comparison, we also show the fixed values of the TIP4P original model\cite{tip4p_water}, which are $d_{\text{OH}} \sim 0.957 \text{ \AA}$ and  $\theta_{\text{HOH}} \sim 104.52 ^\circ$. The spread in the $d_{\text{OH}}$ and $\theta_{\text{HOH}}$
is particularly not captured in the narrowest nanotube ($d \sim 9.5 \text{ \AA}$), potentially explaining the disagreement between the MLP and TIP4P predictions.

\end{appendices}

\end{document}